\documentclass[twocolumn]{aastex62}
\usepackage{enumitem}
\usepackage{graphicx,url}
\usepackage{color}
\usepackage{amsmath}
\usepackage{amssymb}
\usepackage{mathtools}
\usepackage[htt]{hyphenat}

\newcommand*\justify{%
  \fontdimen2\font=0.4em
  \fontdimen3\font=0.2em
  \fontdimen4\font=0.1em
  \fontdimen7\font=0.1em
  \hyphenchar\font=`\-
}

\def\j1550{XTE~J1550$-$564}

\def\rxte{{\it RXTE}}

\def\asca{{\it ASCA}}

\accepted{February 26, 2020}
\submitjournal{ApJ}

\shorttitle{Reflection Spectroscopy of \j1550}
\shortauthors{Connors et al.}

\setlist[itemize]{leftmargin=*}

\begin{document}

\title{Evidence for Returning Disk Radiation in the Black Hole X-ray Binary \j1550}

\correspondingauthor{Riley~M.~T.~Connors}
\email{rconnors@caltech.edu}

\author{Riley~M.~T.~Connors}
\affiliation{Cahill Center for Astronomy and Astrophysics, California Institute of Technology, \\
  Pasadena, CA 91125, USA}
  
  \author{Javier~A.~Garc\'ia}
\affiliation{Cahill Center for Astronomy and Astrophysics, California Institute of Technology, \\
  Pasadena, CA 91125, USA}
  \affiliation{Dr Karl Remeis-Observatory and Erlangen Centre for Astroparticle Physics,\\
  Sternwartstr. 7, D-96049 Bamberg, Germany}
  
  \author{Thomas Dauser}
\affiliation{Dr Karl Remeis-Observatory and Erlangen Centre for Astroparticle Physics,\\
  Sternwartstr. 7, D-96049 Bamberg, Germany}

\author{Victoria Grinberg}
\affiliation{Institut f\"ur Astronomie und Astrophysik, Universität T\"ubingen, Sand 1, 72076 T\"ubingen, Germany}

\author{James~F.~Steiner}
\affiliation{MIT Kavli Institute, 77 Massachusetts Avenue, 37-241, \\
Cambridge, MA 02139, USA}
\affiliation{CfA, 60 Garden St. Cambridge, MA 02138, USA}

\author{Navin Sridhar}
\affiliation{Department of Astronomy, Columbia University,\\
    550 W 120th St, New York, NY 10027, USA}

\author{J\"orn Wilms}
\affiliation{Dr Karl Remeis-Observatory and Erlangen Centre for Astroparticle Physics,\\
  Sternwartstr. 7, D-96049 Bamberg, Germany}

\author{John Tomsick}
\affiliation{Space Sciences Laboratory, University of California Berkeley,\\
7 Gauss Way, Berkeley, CA 94720-7450}

\author{Fiona Harrison}
 \affiliation{Cahill Center for Astronomy and Astrophysics, California Institute of Technology, \\
  Pasadena, CA 91125, USA}
  
  \author{Stefan Licklederer}
  \affiliation{Dr Karl Remeis-Observatory and Erlangen Centre for Astroparticle Physics,\\
  Sternwartstr. 7, D-96049 Bamberg, Germany}








\begin{abstract}

 We explore the accretion properties of the black hole X-ray binary \j1550\ during its outbursts in 1998/99 and 2000. We model the disk, corona, and reflection components of X-ray spectra taken with the {\it Rossi X-ray Timing Explorer} (\rxte), using the {\tt relxill} suite of reflection models. The key result of our modeling is that the reflection spectrum in the very soft state is best explained by disk self-irradiation, i.e., photons from the inner disk are bent by the strong gravity of the black hole, and reflected off the disk surface. This is the first known detection of thermal disk radiation reflecting off the inner disk. There is also an apparent absorption line at $\sim6.9$~keV which may be evidence of an ionized disk wind. The coronal electron temperature ($kT_{\rm e}$) is, as expected, lower in the brighter outburst of 1998/99, explained qualitatively by more efficient coronal cooling due to irradiating disk photons. The disk inner radius is consistent with being within a few times the innermost stable circular orbit (ISCO) throughout the bright-hard-to-soft states (10s of $r_{\rm g}$ in gravitational units). The disk inclination is low during the hard state, disagreeing with the binary inclination value, and very close to $90^{\circ}$ in the soft state, recovering to a lower value when adopting a blackbody spectrum as the irradiating continuum.

\end{abstract}

\keywords{accretion, accretion disks -- atomic processes -- black hole physics}

%
%
%
\section{Introduction}\label{sec:intro}

The study of accretion as a physical process has provided us with a myriad of interesting conclusions regarding the nature of black holes (BHs) and strong gravitational fields. This is largely due to the capabilities we have to approach the topic across vast variability timescales, distances, and scale sizes. Active galactic nuclei (AGN), due to the linear relation between black hole mass and dynamical timescale, are not observed to evolve significantly on human timescales---with the exception of a few newly discovered changing-look quasars, e.g., \citealt{McElroy2016,Yang2018}. However, their smaller cousins, black hole X-ray binaries (BHBs), exhibit high variations in flux and spectral shape over just days to weeks (see, e.g., \citealt{Nowak1995,Homan2005,rm06}). As such, in depth modeling of BHBs as they evolve during outbursts allows us to understand the driving physical conditions for observable changes, and to attempt to relate this understanding to the behavior of AGN. 

Many such studies of BHB spectral evolution have been conducted. The broadly classified ``hard" and ``soft" states are now mostly understood to be the result of combinations of several principal components: thermal blackbody emission from a multitemperature accretion disk \citep{SS1973, Done2007}; a hard, power-law component, originating from an optically thin gas which inverse-Compton (IC) scatters the thermal disk photons, and is either a hot compact corona \citep{Haardt1993,Dove1997}, or sits in the base of a relativistic jet \citep{mnw05}; and a reflected component of emission, which we expect is generated by the power-law emission illuminating the accretion disk \citep{Fabian1989,Garcia2014}. 

\j1550 is a Galactic, transient BHB, first detected by the All-Sky Monitor on board the {\it Rossi X-ray Timing Explorer} (\rxte) on September 6 1998 \citep{Smith1998}. Subsequent daily monitoring with \rxte\ for the following eight months \citep{Sobczak2000} revealed a significant 7-Crab X-ray flare just two weeks into the outburst. The dynamical characteristics of \j1550\ are well-determined. Optical/Infrared observations made with the 6.5-meter Magellan telescopes have led to strong constraints on the BH mass, source distance, orbital period, and binary inclination: $M_\mathrm{BH}=9.1 \pm 0.6~M_\odot$, $D=4.4^{+0.6}_{-0.4}~\mathrm{kpc}$, $P_\mathrm{orb}=1.54~\mathrm{days}$ and $i=75^{\circ} \pm 4^{\circ}$ \citep{Orosz2002, Orosz2011}. Additionally, X-ray timing studies of the initial outburst in 1998/99 with \rxte\ revealed quasi-periodic oscillations (QPOs) throughout the outburst \citep{Remillard2002a}. \j1550\ has since gone into outburst on four additional occasions, comprising one full spectral evolution in 2000 \citep{Rodriguez2003}, and three ``failed" outbursts in 2001, 2002, and 2003 (a ``failed'' outburst is one in which the source does not transition from the hard to the soft state; \citealt{rm06}). As such, the X-ray spectral and time variability characteristics of \j1550\ have been extensively studied \citep{Sobczak2000,Homan2001,Remillard2002b, Rodriguez2003, Kubota2004,Dunn2010}.


Several estimates have been made of the dimensionless spin ($a_{\star}=cJ/GM^2$, where $J$ is the spin angular momentum) of the BH in \j1550\ ($0.1\mbox{--}0.9$, \citealt{Davis2006}; $0.76\mbox{--}0.8$, \citealt{Miller2009b}; $0.49^{+0.13}_{-0.20}$, \citealt{Steiner2011}; $0.34\pm0.01$, \citealt{Motta2014}) using either the thermal disk continuum fitting method  \citep{Li2005,McClintock2006},  modeling of relativistic reflection of X-rays off the accretion disk \citep{Ross2005,Ross2007, Brenneman2006}, or modeling of QPOs \citep{Motta2014}. All such modeling, whilst not in perfect agreement quantitatively, reveals the BH spin to be less than maximal, with a rough average value of $a_\mathrm{\star}=0.5$.

There has not yet been a detailed study of relativistic reflection in \j1550\ as the source evolves through its outbursts. We do, however, have a general phenomenological understanding of its hard-to-soft spectral evolution, particularly from the first two outbursts in 1998/99 and 2000 \citep{Sobczak2000,Rodriguez2003}. The hard-to-soft spectral transition during both outbursts is well characterized by a thermal disk component, peaking at $\sim1$~keV in the soft state, and a power-law component which persists through the hard and hard-intermediate states. The power law steepens significantly ($\Gamma\sim 2.5$--$3$) during the intermediate states, typical of the long-known steep power-law states of BHBs (e.g., \citealt{Miyamoto1991,Miyamoto1993}). In addition, curious behavior was found during the 7-Crab flare in the 1998/99 outburst. {\cite{Sobczak2000} modeled the thermal disk spectrum of \j1550, and found that the inner radius of the accretion disk decreases sharply following the flare. However, they do note that this drop in radius could be artificial, i.e., a color correction to the disk spectrum, which is degenerate with disk temperature and radius through the overall flux. 

We previously modeled (\citealt{Connors2019_refl}; from now on C19) the hard-intermediate state broadband ($1$--$200$~keV) X-ray spectrum of \j1550 with the most up-to-date relativistic reflection model, {\tt relxill} \citep{Garcia2014, Dauser2014}. C19 found that \j1550\ appears to have an inner disk inclination of $39^{+0.6}_{-0.4}$~degrees, based on the reflection spectrum, which is $\sim35^{\circ}$ lower than the confirmed binary inclination found by \cite{Orosz2011}. However, this constraint was based on just one simultaneous observation of \j1550\, made with the {\it Advanced Satellite for Cosmology and Astrophysics} (\asca) and \rxte\ during the intermediate state. Here, we seek to model the evolution of the reflection spectrum of \j1550\ in order to better characterize the geometry and thermal properties of its inner accretion disk and corona. 

In this paper we explore the disk, coronal, and reflection properties of \j1550\ during its first two complete outbursts in 1998/99 and 2000, by physically modeling a sample of archival \rxte\ observations. In Section~\ref{sec:data} we describe the \rxte\ data reduction process. In Section~\ref{sec:modeling} we outline our spectral modeling strategy and procedure, and detail the results. In Section~\ref{sec:discussion} we detail the implications of our reflection modeling results, and in Section~\ref{sec:conclusion} we give a concluding summary. The most striking result of our modeling, as discussed in Section~\ref{sec:return}, is that we find the best-fit spectral reflection model during the very soft state of \j1550 is produced by an irradiating blackbody continuum; we have found evidence for emission returning from the inner disk onto itself due to the strong gravity of the BH. 
%
%
%
 \section{\rxte\ ~Data reduction}
 \label{sec:data}
 
 \rxte\ ~observed \j1550 over 400 times, with more than half of these observations taken during the first outburst in 1998/99. All the data from these observations is publicly available on the \rxte\ archive via the HEASARC (High Energy Astrophysics Science Archive Research Center). We extracted data from the Proportional Counter Array (PCA) lying within 10~min of the South Atlantic Anomaly (SAA). Since proportional counter unit (PCU) 2 has the best calibration of all the PCUs, and the best coverage (all PCA exposures), we use only the data from this PCU, using all three PCU~2 layers. We then corrected all the PCU~2 spectra using the tool {\tt pcacorr} \citep{Garcia2014b}, and subsequently added 0.1\% systematics to all the PCU~2 channels---these comparatively low systematics are made possible by the reduction in systematic residuals provided by the {\tt pcacorr} tool. The corrections provided by {\tt pcacorr} result from utilizing observations of the Crab with the PCA, and iteratively reducing systematic residuals present in averaged powerlaw fits to a summed Crab spectrum. We refer the reader to \cite{Garcia2014b} for the details of this correction, and just note here that for PCA spectra composed of $\gtrsim10^7$~counts, there is up to an order of magnitude increase in sensitivity to faint spectral features. We then group the PCU~2 spectra at a signal-to-noise of 4 based upon visual inspection of the faintest spectra and their backgrounds, such that there are sufficient counts per bin up to high energies ($> 20$). We restrict our spectral fitting to $3\mbox{--}45$~keV.

 \begin{figure*}[!ht]
 \centering
\includegraphics[width=0.8\linewidth]{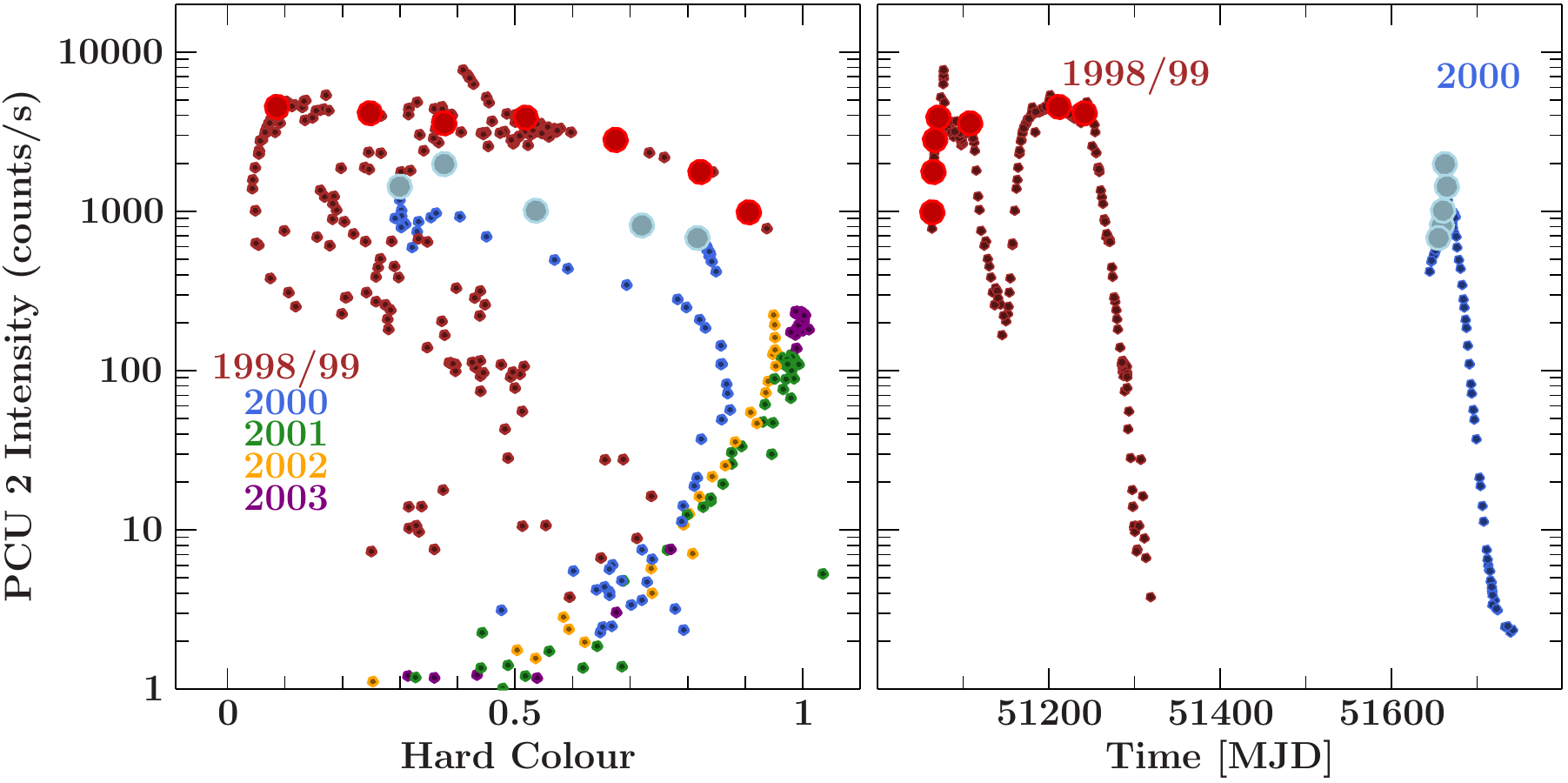}
\caption{{\bf Left}: Hardness-intensity diagram including all \rxte\ observations of \j1550. The hard color is defined as the ratio of source counts in the hard and soft bands, $[8.6$--$18]/[5$--$8.6$]~keV. {\bf Right}: Light curve showing just the first two outbursts of \j1550. Large circles indicate the selected data for this study, seven observations from the first outburst, and five from the second outburst.}
\label{fig:hid}
\end{figure*}

\begin{deluxetable}{lcccr}
\tablecaption{Properties of the selected \rxte-PCA (PCU~2) X-ray spectra from the first and second outbursts of \j1550.}
\tablecolumns{5}
\tablehead{
\colhead{ObsID} & 
\colhead{MJD} & 
\colhead{HR\tablenotemark{a}} & 
\colhead{$N_{\rm counts}$\tablenotemark{b}} & 
\colhead{cts~${\rm s^{-1}}$ \tablenotemark{c}} 
}
\startdata
\hline
& & Outburst 1 & ($10^6$) & \\
\hline
30188-06-03-00 & 51064.0 & 0.91 & $6$ & 986 \\
30188-06-01-01 & 51065.3 & 0.82 & $4$ & 1767 \\
30188-06-04-00 & 51067.3 & 0.67 & $9$ & 2807 \\
30188-06-09-00 & 51071.2 & 0.52 & $13$ & 3873 \\
30191-01-33-00 & 51108.1 & 0.38 & $33$ & 3571 \\
40401-01-50-00 & 51241.8 & 0.25 & $13$ & 4133 \\
40401-01-27-00 & 51211.7 & 0.09 & $12$ & 4525 \\
\hline
& & Outburst 2 & & \\
\hline
50137-02-06-00 & 51654.7 & 0.82 & $2$ & 680 \\
50134-02-01-00 & 51658.6 & 0.72 & $0.7$ & 816 \\
50134-02-01-01 & 51660.1 & 0.54 & $4$ & 1008 \\
50134-02-02-00 & 51662.2 & 0.38 & $2$ & 1978 \\
50134-02-02-01 & 51664.4 & 0.30 & $3$ & 1429 \\
\enddata
\tablenotetext{a}{Hardness ratio given by source counts in [$8.6\mbox{--}18$~keV]/[$5\mbox{--}8.6$~keV] bands.}
\tablenotetext{b}{Number of counts in the $3\mbox{--}45$~keV band of the PCU~2 spectra.}
\tablenotetext{c}{Total $3\mbox{--}45$~keV count rate.}
\tablecomments{Observation 40401-01-50-00 actually follows observation 40401-01-27-00 temporally, but we selected our data in this way to maximize the coverage of spectral hardness. Our sample of observations from outburst 1 covers the period from 8 September 1998 to 4 March 1999. Outburst 2 data covers the period from 20--30 April 2000.} 
\label{tab:data}
\end{deluxetable}

%
%
%
\section{Modeling}
\label{sec:modeling}

We model the changing disk, corona, and reflection components of \j1550\ as it evolves from the hard to soft states during its first two outbursts. Our modeling strategy stems from several key motivating factors:

 1) In C19 we modeled simultaneous \asca\ and \rxte\ observations taken in the hard-intermediate state during the first \j1550 outburst, using the reflection model {\tt relxillCp}. We found that the disk inclination is at $\sim40^{\circ}$, significantly lower than the binary inclination of $\sim75^{\circ}$ \citep{Orosz2011}. Therefore, in this paper we set out to test whether this is true across all spectral states, and whether there is any evolution in the disk inclination. 
 
 2) \cite{Garcia2015} paved the way for global BHB reflection studies using the \rxte\ archive, with the goal of characterizing the disk and coronal parameters of GX~339$-$4, such as disk inner radius, $R_{\rm in}$, and coronal electron temperature, $kT_{\rm e}$, and optical depth, $\tau$. \cite{Garcia2015} found that the inner disk remains within $\sim10~r_{\rm g}$ ($r_{\rm g}=GM/c^2$, where $G$ is the gravitational constant, $M$ is the mass of the BH, and $c$ is the speed of light) during the rise of the hard state, the corona cools, and optical depth increases. However, the focus was on the rise of the hard state, and did not follow the transition from hard to soft toward the outburst peak. We want to model the disk and coronal physics as BHBs transition from the hard to the soft state (similarly to, e.g., \citealt{Sridhar2020}). 
 
 3) \j1550, as shown in Figure~\ref{fig:hid}, shows wide variability in the nature of its outbursts. The initial outburst in 1998/99 was bright, approaching the Eddington limit, double peaked, and reached a very soft spectral state , with a hardness ratio ${\rm (HR)}\sim0.05$. The second outburst in the year 2000 peaked at lower luminosities, and decayed after reaching ${\rm HR}\sim0.3$, so did not become as soft. The following three outbursts were all `failed', remaining spectrally hard and peaking at luminosities a factor of 10 lower. Thus, within the same source we can look for key differences in the accretion physics between outbursts. 
 
 Given these motivators, we selected observations covering the transition from the hard to soft states in outbursts 1 and 2. This selection is shown in Figure~\ref{fig:hid}, highlighted by the large red and blue points. We chose {\it seven} observations from outburst 1, and {\it five} from outburst 2, based on having enough photon statistics to constrain reflection model parameters, and in the case of outburst 2, the availability of data---the source transition rapidly during outburst 2, and thus there are only a few \rxte\ exposures during the hard-to-soft transition. The seven observations taken from outburst 1 span ${\rm HR}=0.09\mbox{--}0.91$, and the five taken from outburst 2 span 0.30\mbox{--}0.82. Table~\ref{tab:data} shows the details of all selected data. 
 
 
Due to the complexity of the data modeling, we only include these 12 observations in the remainder of this paper. We fit all 12 observations with a model including a Comptonized multi-temperature disk blackbody component, relativistically broadened and distant, unbroadened reflection components, and interstellar absorption: {\tt\justify crabcorr * TBabs * (simplcut$\otimes$diskbb + relxillCp + xillverCp)}.

 {\tt Crabcorr} \citep{Steiner2010} corrects the detector response of a given instrument to retrieve the normalization and power-law slope obtained from fits to the Crab spectrum, provided by \cite{Toor1974}. The values adopted by the PCA instrument are $N=1.097$ and $\Delta\Gamma=0.01$. {\tt TBabs} is a model for interstellar absorption using the elemental abundance tables of \cite{Wilms2000}. We use the atomic cross sections of \cite{Verner1996}. 
 
 The model {\tt simplcut} \citep{Steiner2017} is a variant of the model {\tt simpl} \citep{Steiner2009}, and functions as a coronal plasma, inverse-Compton (IC) scattering the disk photons in a convolution kernel. The {\tt simplcut} model includes a coronal electron temperature ($kT_{\rm e}$) and thus contains a high-energy cutoff in the power-law continuum. It is important to be aware of the effects of selecting particular coronal IC continuum components in our modeling. We prefer to use {\tt simplcut} over more physically motivated models, such as {\tt nthComp} \citep{Zdziarski1996,Zycki1999}, because {\tt simplcut} conserves the disk photon flux when calculating the portion of scattered photons, which is set by the parameter $F_{\rm sc}$, with a maximal value of unity resulting in all disk photons being upscattered. As we show explicitly in Section~\ref{sec:refl_results}, since {\tt simplcut} adopts the same spectral shape for the scattered photons as given by {\tt nthComp}, the physical constraints of the corona are identical between the two models. However, since {\tt nthComp} is normalized independently of the disk flux, one can arrive at spurious estimations of the disk flux when modeling hard-state spectra in the $3\mbox{--}45$~keV \rxte\ band. Thus, adopting {\tt simplcut} allows us to constrain the co-evolving disk and corona properly, tracking the inner disk temperature and flux, along with the coronal properties. Similarly, we decide against using the more self-consistent {\tt eqpair} model \citep{Coppi2000}. The {\tt eqpair} model calculates the plasma thermodynamics based upon parameterization of the coronal and disk compactness and coronal optical depth. However, given both that we only model data in the PCA energy band, and need a simple way to relate the coronal properties to the irradiating continuum for reflection, we prefer {\tt simplcut}. In Section~\ref{sec:discussion} we show explicit comparisons of the PCA residuals when applying these different continuum components. The disk photons in our model are provided as a multi-temperature blackbody component {\tt diskbb} \citep{Mitsuda1984}. 
 
 The models {\tt relxillCp} and {\tt xillverCp} are flavours of the {\tt relxill} suite of relativistic reflection models \citep{Dauser2014,Garcia2014}, they are used to calculate the reflection spectrum resulting from the illumination of an IC spectrum atop the accretion disk. {\tt XillverCp} provides the reflection spectrum resulting from this illumination, which produces fluorescent line emission, the most prominent being Fe K emission, as well as Compton down-scattering of higher energy photons, giving the characteristic `{\it Compton hump}'. {\tt RelxillCp} includes the full ray tracing calculations from the irradiating source to the disk and onward to the observer, allowing for a full calculation of the relativistic effects which distort the spectrum, including light-bending effects, Doppler shifts, and gravitational redshifts. 
 
In all our fits we treat the model parameters as follows. The {\tt crabcorr} parameters for offset normalization and photon index are fixed at $N=1.097$ and $\Delta\Gamma=0.01$, respectively. We fix the interstellar absorption hydrogen column density at $N_{\rm H}=10^{22}~{\rm cm^{-2}}$ in accordance with Galactic \ion{H}{1} surveys \citep{Kalberla2005}. Though we found a value of $9.228^{+0.007}_{-0.009}\times10^{21}~{\rm cm^{-2}}$ in C19, in the $3$--$45$~keV band occupied by the PCA data this difference is not impactful on our modeling results, and keeping its value fixed reduces degeneracies. The disk temperature ($T_{\rm in}$) and normalization ($N_{\rm disk}$) in the model component {\tt diskbb} are both kept free. The {\tt simplcut} ReflFrac parameter is fixed to 1, positing only up-scattering in the coronal IC calculation. The photon index of the IC spectrum ($\Gamma$) and electron temperature ($kT_{\rm e}$) are both kept free. We fix the black hole spin to $a_{\star}=0.5$ in rough accordance with the previous spectral continuum fitting, reflection fitting, and time-variability modeling results for \j1550\ \citep{Davis2006,Miller2009b,Steiner2011,Motta2014}. We fix the emissivity index for the illumination of the disk to $q=3$ throughout the disk, since the emissivity profile is typically shallow for non-maximal BH spin \citep{Dauser2013}. The reflection fraction is fixed to -1 such that the reflection components of {\tt relxillCp} and {\tt xillverCp} exclude the illuminating continuum, already provided by {\tt simplcut$\otimes$diskbb}. The photon index ($\Gamma$) and electron temperatures ($kT_{\rm e}$) are tied to the corresponding values in {\tt simplcut}. The disk inclination ($i$) and iron abundance ($A_{\rm Fe}$) are all left as free parameters, and tied between the {\tt relxillCp} and {\tt xillverCp} models. The disk ionization ($\log\xi$) is left to vary freely in the {\tt relxillCp} component, and fixed at $\log\xi=0$ in the {\tt xillverCp} component, representing distant, near-neutral reflection. The inner-disk radius in the {\tt relxillCp} component, $R_{\rm in}$, is left free, and influences the relativistic effects as calculated in the model. The {\tt xillverCp} and {\tt relxillCp} components are normalized independently. 
 
 In the following sections (\ref{sec:hardtosoft}, \ref{sec:vsoft}, \ref{sec:refl_results}), we begin by showing some results of phenomenological fits to our selected data, move on to a discussion of interesting features detected in the very soft state, and then show the full results of our relativistic reflection modeling as discussed in this section. 
  
 \begin{figure}[h]
\centering
\includegraphics[width=\linewidth]{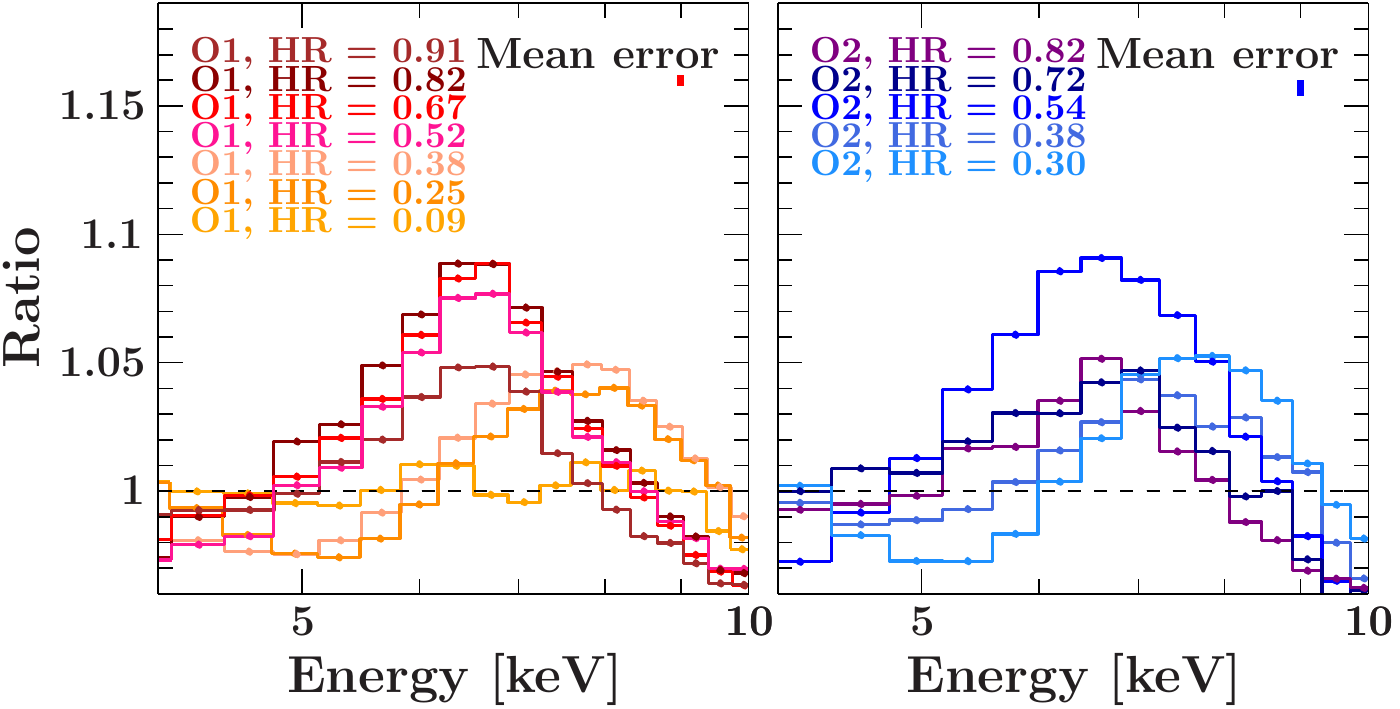}
\caption{Fe K line ratios after fitting the spectral continuum model {\tt TBabs*simplcut$\otimes$diskbb} to all our selected data. The left panel shows the ratio residuals for outburst 1 data, and the right for outburst 2. Error bars have been removed from the residuals for clarity, the average total $\pm$ errors are shown in the top right of each panel. }
\label{fig:line}
\end{figure}

\subsection{Hard-to-soft transition}
\label{sec:hardtosoft}

Figure~\ref{fig:line} shows the evolution of data residuals when fitting the model {\tt\justify TBabs*(simplcut$\otimes$diskbb)} to the PCU~2 spectra in our selected sample. The goal of fitting such a model is to isolate the Fe K emission and edge features.  

It is not possible to definitively quantify a shift in the centroid energy of the Fe K line, due to the limited energy resolution of the PCA detector ($\sim1$~keV at $6$~keV). However, we see more blueward line emission as \j1550\ transitions to the soft state. This is particularly pronounced in outburst 1, during which time the source is brighter. The reasons for this evolution are not clear, but it could possibly be due either to geometrical changes in the inner flow, i.e., the disk inclination may be varying, or alternatively the result of distinct changes in the irradiating spectrum. It is also possible that we are seeing excess emission in the $7$--$9$~keV band that need not necessarily be associated with the Fe K reflected emission. 

\subsection{The very soft state: additional features}
\label{sec:vsoft}

In the very soft state of \j1550, represented in our selected sample by observation 40401-01-27-00, there are prominent features in both the $4$--$5$~keV band, and at $\sim6.8$--$7$~keV (see Figure~\ref{fig:softstate}). In order to explore these, we took a more comprehensive look at the multiple observations taken during this soft branch (${\rm HR}<0.1$) by selecting 11 PCA spectra within an observation window of $\sim13$~days during the secondary rise of the 1998/99 outburst (${\rm HR}=0.09$).

\begin{figure}[h]
\centering
\includegraphics[width=\linewidth]{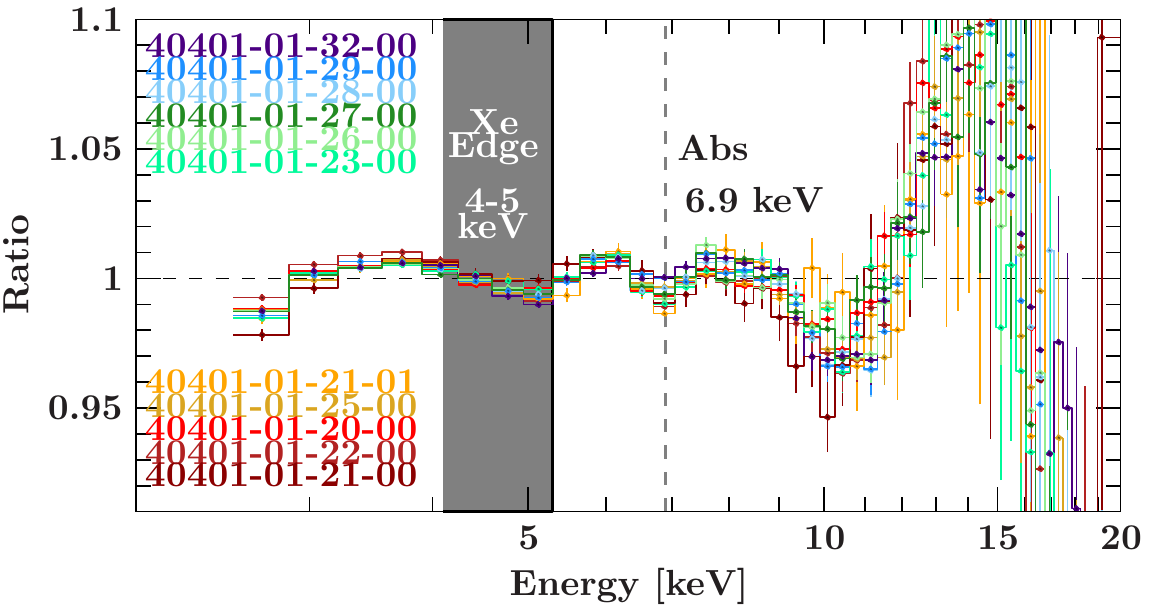}
\caption{Ratio residuals remaining after fitting a basic spectral model, {\tt TBabs*simpl$\otimes$diskbb}, to data within the soft branch of the 1998/99 outburst. All 11 selected spectra show similar features in the 4--5~keV band, and at $\sim6.9$~keV. }
\label{fig:softstate}
\end{figure}

Figure~\ref{fig:softstate} shows curious features in the PCU~2 spectra of 11 individual observations. A striking and unexpected absorption signature appears at $\sim6.8$--$7$~keV. This feature has not been reported in previously analysed PCA data of \j1550 during the 1998/99 outburst \citep{Sobczak2000}, nor in any other observations of the source. The reasons for this are likely that \cite{Sobczak2000} necessarily added 0.5\% systematics to the PCA channels in their analysis, undoubtedly masking this feature. We refer the reader to \cite{Garcia2014} for details of the {\tt pcacorr} tool, showing the complex systematics the tool removes (see also Appendix~\ref{sec:app}). Since we were able to reduce many of the PCA systematics using the {\tt pcacorr} tool, and thus add only 0.1\% systematics, this feature may now have become observable. The explanations for the feature are unclear, but could be evidence of either of the following: (i) an absorption line from an outflowing disk wind, or (ii) a feature inherent to the PCA detector. 

Disk winds are ubiquitous in BHB soft states \citep{Ponti2012}, thus it is not unexpected that we may see such signatures, though they have not previously been detected in \j1550. If present in a wind, this feature is likely to coincide with the \ion{Fe}{26} line, previously found in BHBs in the soft state (e.g., \citealt{Lee2002,Miller2006a}). Thus, to test the validity of the claim that we may be seeing the same feature in our PCU~2 data in the soft state, we performed full phenomenological fits to the softest observation in our sample.

We fit observation 40401-01-27-00 (${\rm HR} = 0.09$) using the model {\tt\justify [crabcorr * TBabs * smedge(simplcut $\otimes$ diskbb + gau + gau) * edge]}. The first Gaussian component represents the Fe K emission line due to reflection, and the second Gaussian has negative normalization to represent the \ion{Fe}{26} absorption line from the disk wind. The energies of the emission and absorption line are fixed at 6.4~keV and 6.9~keV respectively. The width of the absorption line is fixed at $\sigma=0.01$~keV, but we allow the emission line width to vary freely such as to represent relativistic smearing at the inner disk. The {\tt smedge} component represents the relativistically smeared iron edge (Ebisawa PhD thesis, implemented by Frank Marshall). We fix the edge width to 7.1~keV, allow the edge energy to vary between $7\mbox{--}9$~keV, and the optical depth $\tau$ to vary freely. The {\tt edge} component is included at $\sim4.8$~keV, representing the xenon L edge in the PCU~2 layers. Figure~\ref{fig:gauabs} shows the resultant fit, achieving $\chi^2_{\nu}=64/39=1.6$. The equivalent width (EW) of the absorption line is $\sim33$~eV, comparable to those found for the \ion{Fe}{26} line in other soft-state BHBs (e.g., \citealt{Miller2006a}; $\sim40$~eV). 


\begin{figure}[h]
\centering
\includegraphics[width=\linewidth]{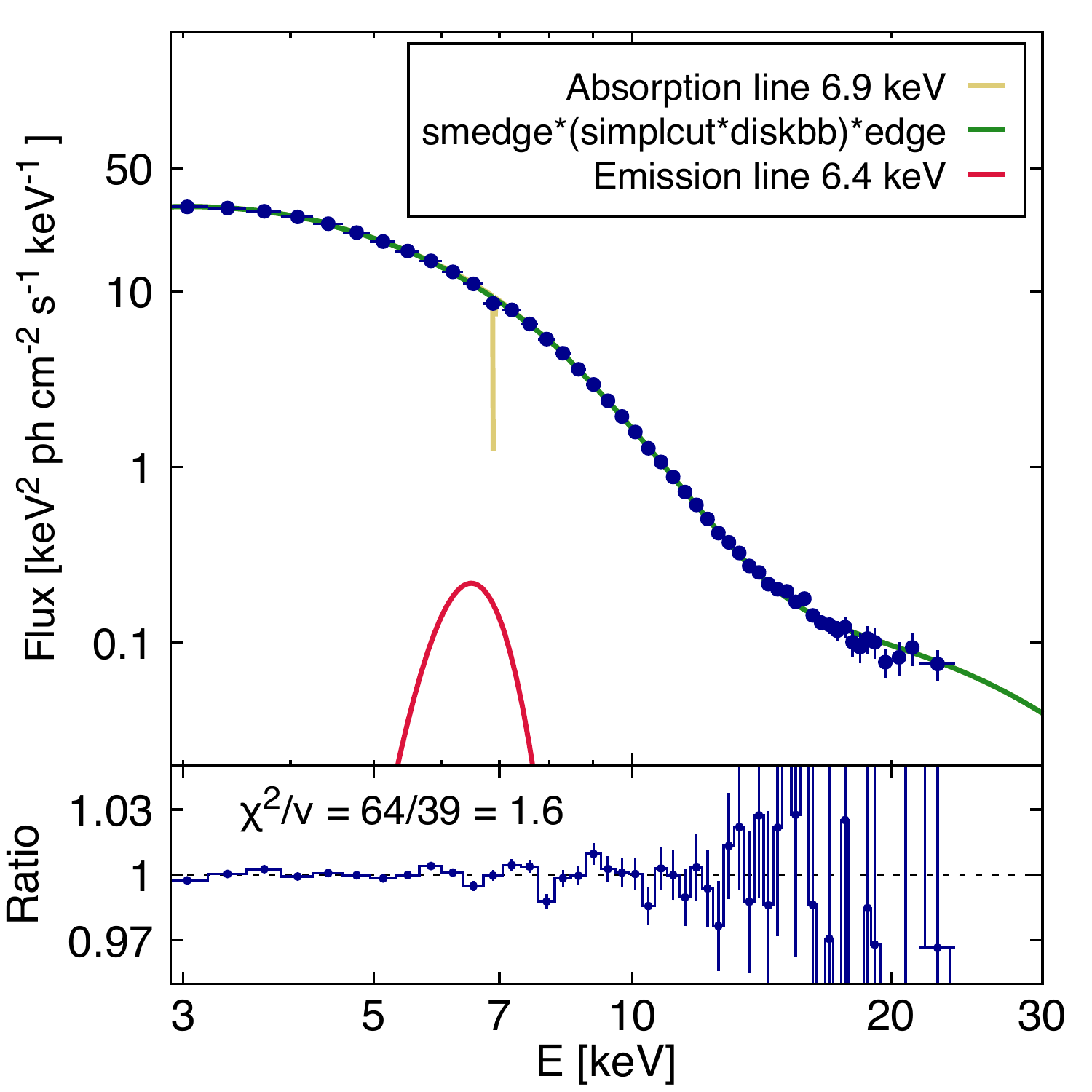}
\caption{Fit of model {\tt\justify crabcorr * TBabs * smedge (simplcut$\otimes$diskbb + gau + gau) * edge} to 40401-01-27-00. The first Gaussian component represents the broad Fe K emission line due to reflection. The second Gaussian component represents absorption in a disk wind, likely \ion{Fe}{26}. We fix the emission line at 6.4~keV and allow the width to vary freely. The absorption line is fixed at 6.9~keV, with $\sigma=0.01$~keV. The bottom panel shows the data-to-model ratios. }
\label{fig:gauabs}
\end{figure}

However, since this component had been revealed to us after applying the {\tt pcacorr} tool to the PCA data, we cannot rule out the possibility that this feature is inherent to the PCU~2 detector. Given the softness of the data, and thus number of X-ray counts in the low-energy PCA channels, absorption features can manifest where they were previously left undetected. This was noted by \cite{Garcia2015} in their global study of GX~339$-$4. Two apparent absorption features were detected in the PCU~2 spectra of GX~339$-$4 at $\sim5.6$~keV and $\sim7.2$~keV. \cite{Garcia2015} proposed that these could have appeared due to the uncertain energy resolution of the PCA. However, in our data we only detect an absorption feature at $\sim6.9$keV, and it only appears during the soft state, whereas \cite{Garcia2015} detected both features in the bright hard state of GX~339$-$4. 

The edge at $\sim4$--$5$~keV has been previously reported in \rxte\ observations of bright sources, and has also been discussed in the relevant calibration papers \citep{Jahoda_2006a,Shaposhnikov2012}. It is thus well-known that the Xe L-edge region still requires modeling, because this feature is not fully accounted for in the calibration. 

In the following Section we discuss the results of full reflection modeling of our selected sample of data, and include the additional features discussed here, applying an edge component to represent xenon from the detector (necessary in observations exceeding $\sim10^7$~counts, with significant disk emission, i.e., soft) wherever it is needed by the data, and a Gaussian absorption feature at $\sim6.9$~keV to model out the residual feature around the Fe K line.

\subsection{Reflection modeling results}
\label{sec:refl_results}

\begin{figure*}
\centering
\includegraphics[width=\linewidth]{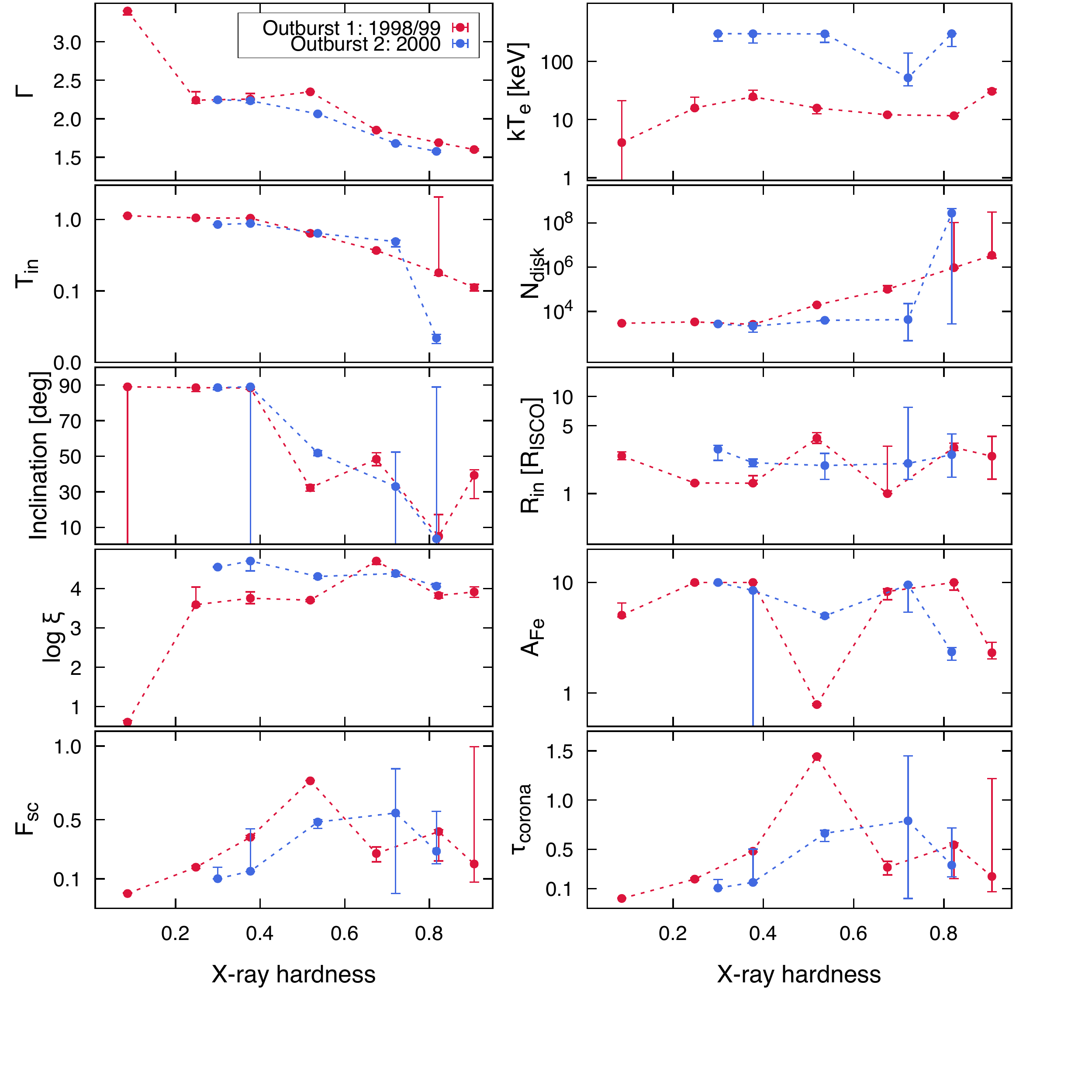}
\vspace{-50pt}
\caption{Parameters and their uncertainties against spectral hardness. All data were fit with the model {\tt\justify crabcorr * TBabs * (simplcut$\otimes$diskbb + relxillCp + xillverCp)} with the following exceptions: in cases in which a xenon edge is required in the $4$--$5$~keV band, and in the very soft state of outburst 1, where an absorption line commensurate with a disk wind is required at 6.9~keV. Red points show the parameter trends for outburst 1, and blue for outburst 2. The coronal optical depth is calculated as $\tau_{\rm corona}=-\ln(1-F_{\rm sc})$.}
\label{fig:pars}
\end{figure*}

\begin{figure*}
\centering
\includegraphics[width=0.48\linewidth]{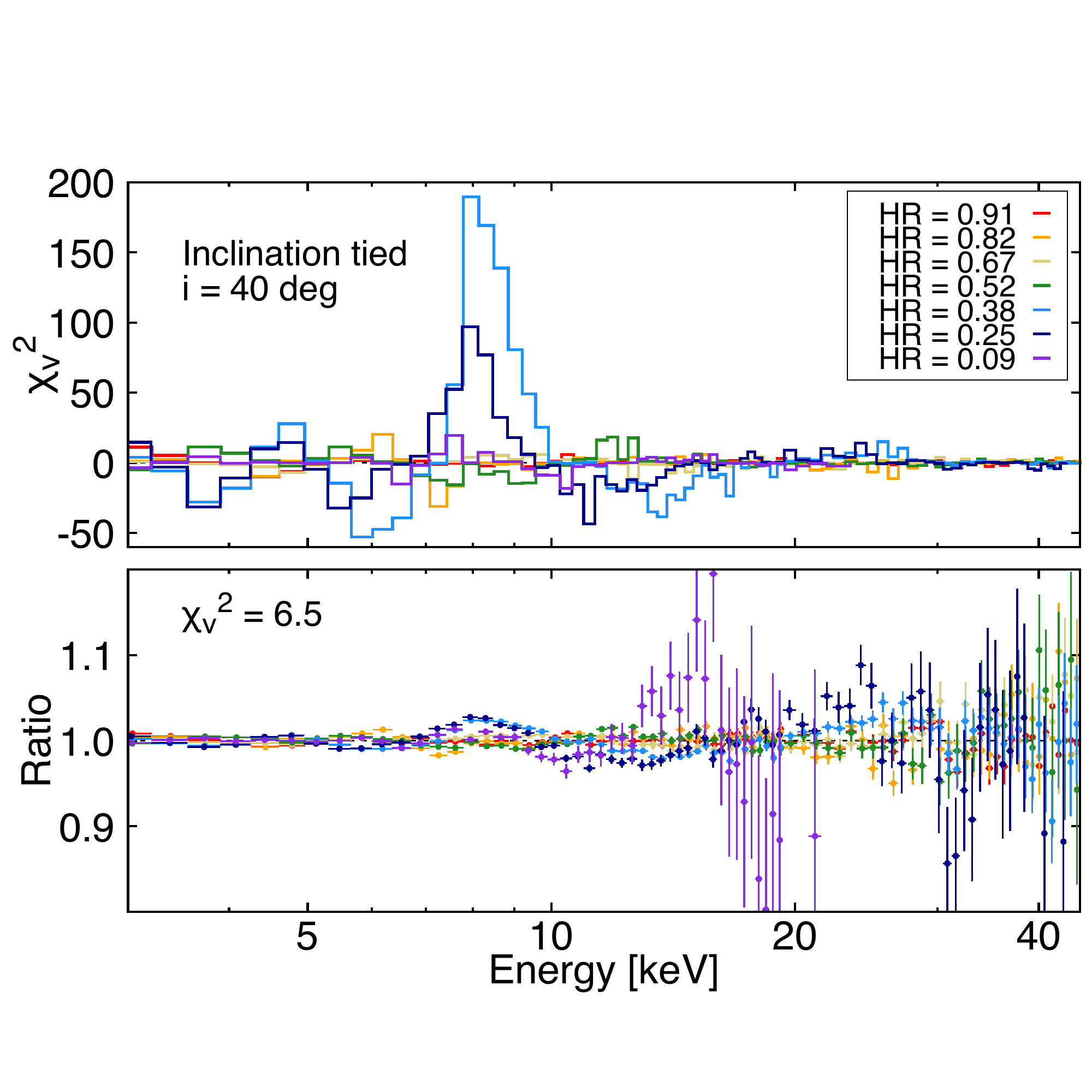}
\includegraphics[width=0.48\linewidth]{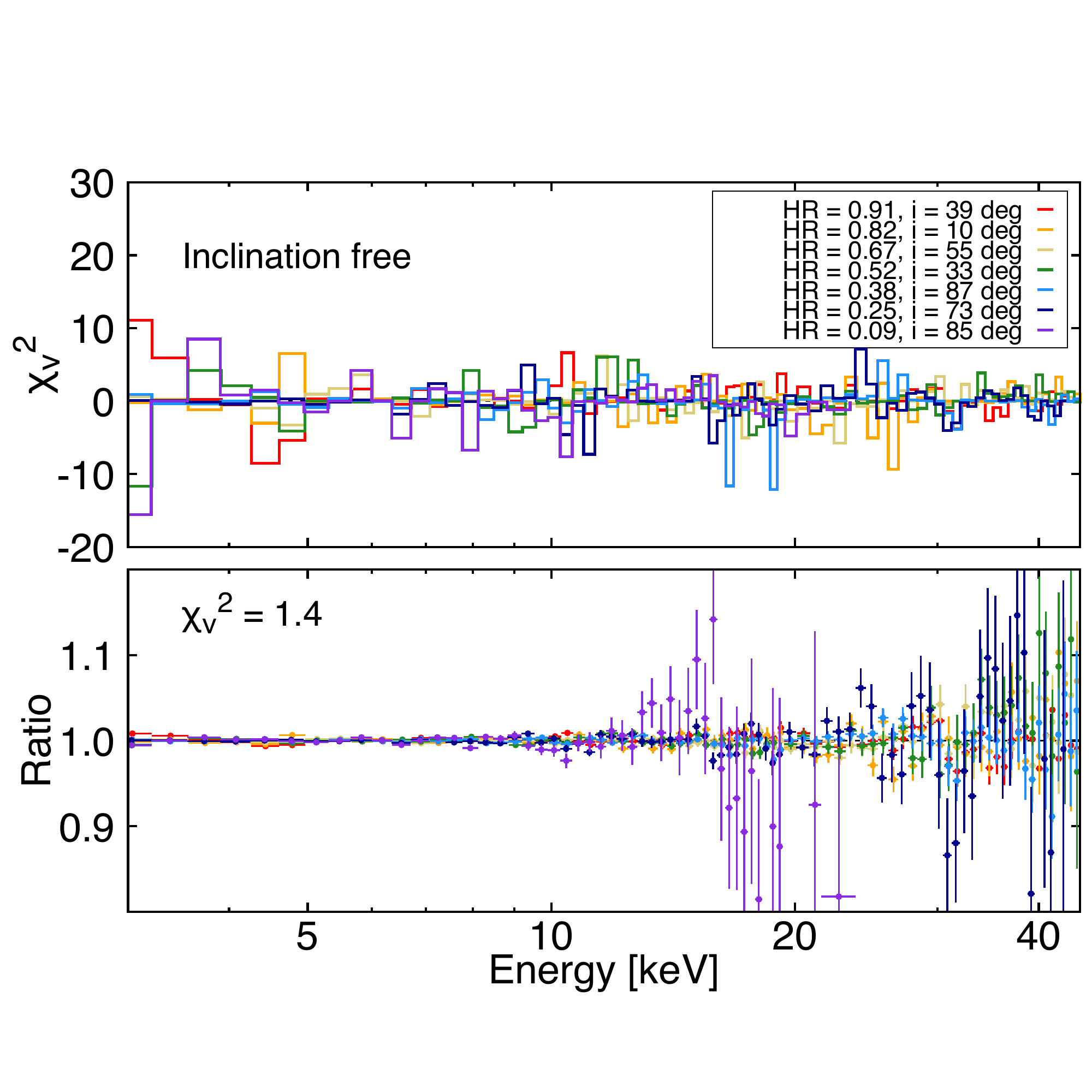}
\vspace{-20pt}
\caption{$\chi^2$ and ratio residuals resulting from joint fits to our selected sample of PCA spectra from outburst 1 of \j1550. The iron abundance, $A_{\rm Fe}$, is tied during both sets of joint fits. The left panel shows the result of tying the disk inclination across all spectral fits, the right panel shows the vast improvement achieved with the inclination allowed to vary between spectral fits.}
\label{fig:incl}
\end{figure*}

\begin{figure}[h]
\centering
\includegraphics[width=\linewidth]{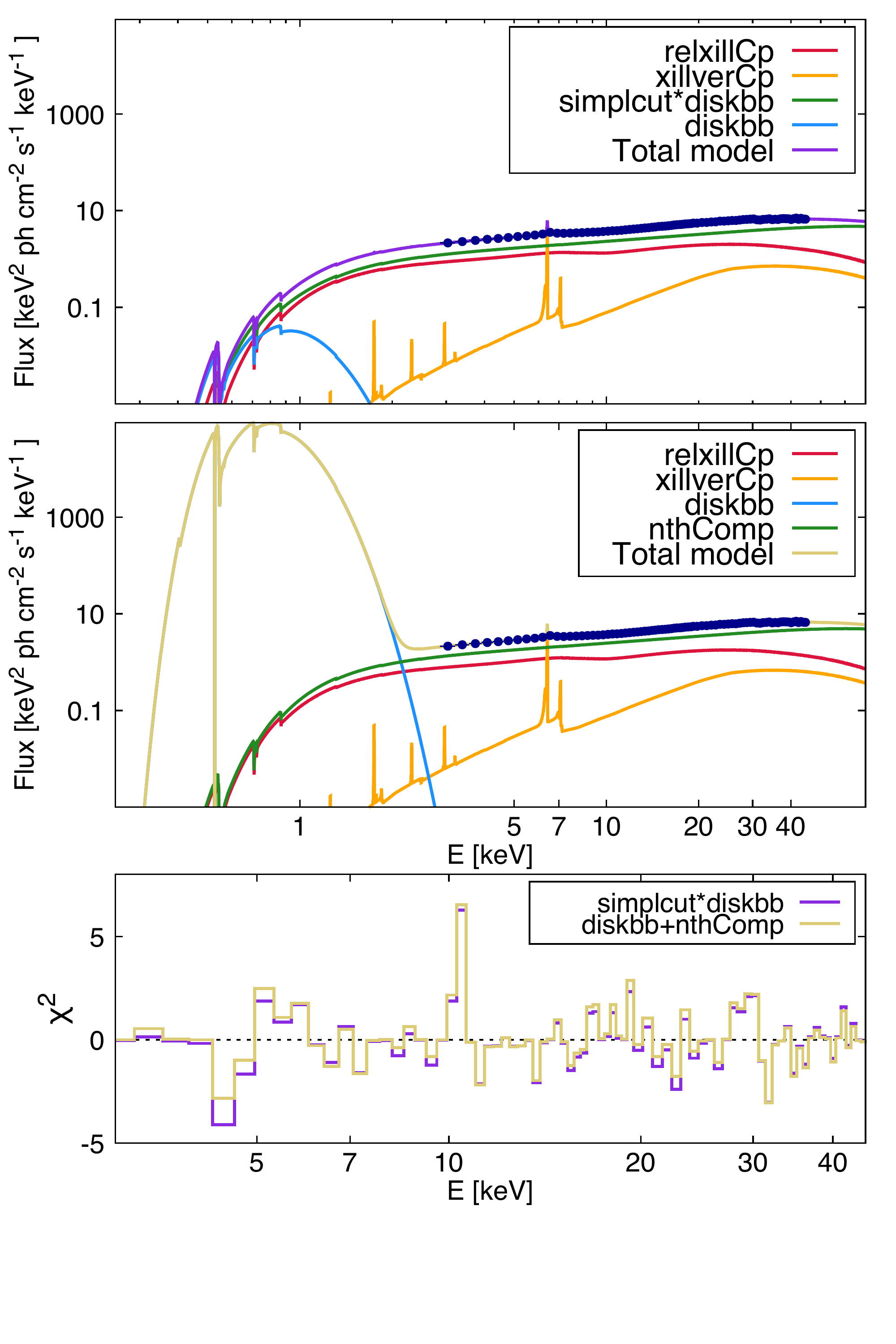}
\vspace{-50pt}
\caption{The PCU~2 spectrum of \j1550\ in the hard state (ObsID 30188-06-03-00) fit with two flavors of disk, corona, and reflection models. In the top panel, the coronal IC spectrum is described by the convolution {\tt simplcut$\otimes$diskbb}, in the middle panel by {\tt nthComp}. The bottom panel then shows the overall $\chi^2$ residuals of the two fits.}
\label{fig:simpl_v_nthComp}
\end{figure}

We have established that there are complex features in the Fe region, particularly in the very soft branch of outburst 1. Therefore, we now show results of our best reflection models applied to these data, taking into account the complex residual features as already described. Figure~\ref{fig:pars} shows the key reflection modeling parameters and their associated uncertainties as a function of X-ray hardness, allowing direct comparison between outbursts 1 and 2. Tables~\ref{tab:params1} and \ref{tab:params2} show the numerical values corresponding to Figure~\ref{fig:pars}, along with the other model parameters. There are several interesting results to notice.

Firstly, we see relatively consistent evolution of the thin accretion disk properties between the two outbursts. The inner disk temperature ($T_{\rm in}$) and normalization ($N_{\rm disk}$) increase and decrease respectively as the source transitions from the hard to the soft state. This is consistent with the inner disk moving closer to the innermost stable circular orbit (ISCO). However, constraints on $R_{\rm in}$ from the reflection component are not very strong, but largely consistent with being, if not at the ISCO, within a factor of a few. The ISCO, for a prograde BH spinning at $a_{\star}=0.5$, is at $4.23~r_{\rm g}$, thus the range of disk inner radii from the hard to soft states is from a maximum of $\sim18~r_{\rm g}$ and $\sim34~r_{\rm g}$ in outbursts 1 and 2 respectively, down to $4.23~r_{\rm g}$. We did not relate the $R_{\rm in}$ parameter of the reflection model {\tt relxillCp} to the disk normalization ($N_{\rm disk}=(R_{\rm in}/D_{10})^2\cos\theta$, where $D_{10}$ is the distance to the source in units of 10~kpc, and $\theta$ is the disk inclination) in our modeling. However, the estimates of $R_{\rm in}$ derived from the $N_{\rm disk}$ constraints broadly agree with the reflection modeling results, with the exception of those at ${\rm HR}>0.8$, i.e., the bright hard state. However, this calculation does not take into account the uncertainty in the color temperature of the inner disk (which can be up to a factor of 2; see, e.g., \citealt{Davis2005}).


Secondly, the properties of the Comptonizing plasma show very similar behavior to that observed in previous global reflection studies of GX~339$-$4 \citep{Garcia2015}. The corona remains much hotter during outburst 2, the fainter outburst, whereas the photon index ($\Gamma$) of the coronal spectrum is almost identical throughout the transition. However, closer inspection of the coronal temperature constraints ($kT_{\rm e}$) during outburst 2 reveals that we mostly only achieve lower limits, and those lower limits are typically far beyond the maximum energy of the PCA ($>45$~keV). Therefore, in order to ensure we have not limited ourselves by the exclusion of the available HEXTE data in this case, we re-modeled those observations (outburst 2) with the HEXTE data included. We selected the HEXTE cluster A and B data, including data between $20\mbox{--}200$~keV. HEXTE B spectra were corrected using the {\tt hexBcorr} tool \citep{Garcia2016}, and we grouped both HEXTE A and B spectra by factors of 2, 3, and 4 in the $20\mbox{--}30$, $30\mbox{--}40$, and $40\mbox{--}250$~keV ranges respectively, in order to achieve an oversampling of $\sim3$ times the instrumental resolution. We then grouped all HEXTE spectra by a signal-to-noise ratio of 4, just as we did with the PCA, in order to achieve the required statistics per bin. We fit the PCA and HEXTE A/B data simultaneously, adopting free normalisation constants in the {\tt crabcorr} model to account for cross-calibration between instruments. The results are shown in Table~\ref{tab:params2_2}. The coronal electron temperature, $kT_{\rm e}$, remains very high, and can only be constrained in the first two observations. Here the values ($80^{+70}_{-30}$~keV and $70^{+10}_{-10}$~keV) are still significantly higher than those found for outburst 1, confirming our result that the corona is hotter during outburst 2. In addition, other key reflection properties do not differ significantly, though we do find the inclusion of HEXTE data does allow for slightly tighter constraints on the disk inclination in the hard state; as in outburst 1, we find a low value, $40^{+10}_{-10}$ degrees.

Thirdly, with some variations, the disk inclination evolves from low to high as the source transitions from the hard to the soft state. In both outbursts 1 and 2, below ${\rm HR}=0.4$ we derive very high (almost maximal) disk inclination from reflection modeling, although we find it is unconstrained in the softest observation (40401-01-27-00), and fail to find a good fit to the data. To verify that the variable inclination constraints obtained from our individual modeling are not driven by modeling degeneracies, we fit jointly to all the outburst 1 spectra with the iron abundance (which we do not expect to vary significantly in an outbursting disk) and disk inclination tied between each spectral model. The results of this test are shown in Figure~\ref{fig:incl}. One can see that when attempting to tie the disk inclination across all the fits, we cannot achieve a good enough fit to the data, and we can see strong residual features around the Fe K emission line. When we allow the inclination to vary freely, we can achieve a reasonable joint fit to all our data, and the inclinations settle to values similar to those found via individual fits (Table~\ref{tab:params1}). There is one additional possible explanation for the apparent blueward shifting of the broad Fe K line during the transition to the soft state: there could be a progressive drop in the narrow line component. Since the narrow component, modeled here by {\tt xillverCp}, is a near-neutral reflector, the line centroid is naturally at lower energies than the broad component (which is more ionized). The decrease in strength of the narrow component could result in an apparent blueward shift in the line profile in the PCA data, given the low spectral resolution. The inherent model degeneracies here, as well as the ubiquitous weakness of the narrow line component, mean we cannot conclusively test this idea. 

It is possible that the geometry of the inner accretion flow may be evolving during outburst, which we discuss in Section~\ref{sec:discussion}, and in detail in C19. However, it is curious that a model predicting high levels of illumination of the disk by IC, power-law-like emission, should suddenly yield wildly different disk inclinations when applied in the soft state; we could be misrepresenting the spectral shape of the illumination. We address this further in the discussion (Section~\ref{sec:discussion}).

 As a cross-check that our constraints on the evolving disk and coronal parameters are not skewed by the coronal IC continuum we selected ({\tt simplcut}), we compare a fit to our hardest sample spectrum with the {\tt simplcut} and {\tt nthComp} models. Figure~\ref{fig:simpl_v_nthComp} shows the two models as unfolded spectra, along with the data and the $\chi^2$ residuals. The coronal parameters of these two fits are statistically indistinguishable, but there is an alarming discrepancy in the disk properties. When applying {\tt nthComp} in the hard state, the {\tt diskbb} component is normalized independently from the corona. As such, due to the lack of data coverage below $\sim3$~keV, the disk is artificially pushed to low temperatures and very high flux. Whereas when we apply the convolution {\tt simplcut$\otimes$diskbb}, the disk flux is constrained by the coronal flux. This shows the motivation for applying the {\tt simplcut$\otimes$diskbb} convolution model to represent the co-evolving disk and corona in our modeling.



\def\oIOnega{$1.60^{+0.01}_{-0.02}$}
\def\oIOnefsc{$0.2^{+0.8}_{-0.1}$}
\def\oIOnekTe{$30^{+3}_{-2}$}
\def\oIOneTin{$0.11^{+0.01}_{-0.01}$}
\def\oIOnendisk{$3.4_{-0.9}^{+300}\times10^6$}
\def\oIOneq{$0.1$}
\def\oIOneincl{$39^{+3}_{-13}$}
\def\oIOneRin{$<3$}
\def\oIOnelogxi{$3.9^{+0.1}_{-0.1}$}
\def\oIOneAfe{$2.3^{+0.6}_{-0.3}$}
\def\oIOnenrel{$5^{+1}_{-2}$}
\def\oIOnenxil{$6^{+1}_{-2}$}
\def\oIOnechi{$70$}
\def\oIOnenu{$69$}
\def\oIOnechired{$1.01$}

\def\oITwoga{$1.689^{+0.002}_{-0.001}$}
\def\oITwofsc{$0.42^{+0.01}_{-0.20}$}
\def\oITwokTe{$11.6^{+0.2}_{-0.2}$}
\def\oITwoTin{$0.18^{+1.89}_{-0.01}$}
\def\oITwondisk{$0.9^{+102}_{-0.5}\times10^6$}
\def\oITwoq{$0.1$}
\def\oITwoincl{$<17$}
\def\oITwoRin{$3.0^{+0.3}_{-0.2}$}
\def\oITwologxi{$3.83^{+0.06}_{-0.07}$}
\def\oITwoAfe{$>9$}
\def\oITwonrel{$2.65^{+0.20}_{-0.05}$}
\def\oITwonxil{$<1$}
\def\oITwochi{$88$}
\def\oITwonu{$69$}
\def\oITwochired{$1.27$}

\def\oIThreega{$1.852^{+0.008}_{-0.003}$}
\def\oIThreefsc{$0.27^{+0.04}_{-0.06}$}
\def\oIThreekTe{$12.1^{+0.2}_{-0.2}$}
\def\oIThreeTin{$0.367^{+0.003}_{-0.021}$}
\def\oIThreendisk{$10^{+5}_{-2}\times10^5$}
\def\oIThreeq{$0.1$}
\def\oIThreeincl{$48^{+4}_{-4}$}
\def\oIThreeRin{$<3$}
\def\oIThreelogxi{$>4.61$}
\def\oIThreeAfe{$8.3^{+0.5}_{-1.3}$}
\def\oIThreenrel{$16^{+1}_{-1}$}
\def\oIThreenxil{$12^{+4}_{-1}$}
\def\oIThreechi{$68$}
\def\oIThreenu{$69$}
\def\oIThreechired{$0.98$}

\def\oIFourga{$2.350^{+0.001}_{-0.001}$}
\def\oIFourfsc{$0.764^{+0.003}_{-0}$}
\def\oIFourkTe{$15.7^{+0.2}_{-3.2}$}
\def\oIFourTin{$0.637^{+0.002}_{-0.003}$}
\def\oIFourndisk{$1.955^{+0.008}_{-0.009}\times10^5$}
\def\oIFourq{$0.1$}
\def\oIFourincl{$32^{+2}_{-2}$}
\def\oIFourRin{$3.7^{+0.5}_{-0.4}$}
\def\oIFourlogxi{$3.71^{+0.01}_{-0.04}$}
\def\oIFourAfe{$0.79^{+0.02}_{-0.02}$}
\def\oIFournrel{$19.8^{+0.4}_{-0.4}$}
\def\oIFournxil{$43^{+1}_{-1}$}
\def\oIFourchi{$66$}
\def\oIFournu{$69$}
\def\oIFourchired{$0.95$}

\def\oIFivega{$2.256^{+0.073}_{-0.007}$}
\def\oIFivefsc{$0.382^{+0.012}_{-0.004}$}
\def\oIFivekTe{$25^{+7}_{-2}$}
\def\oIFiveTin{$1.040^{+0.008}_{-0.003}$}
\def\oIFivendisk{$2.61^{+0.09}_{-0.06}\times10^4$}
\def\oIFiveq{$0.1$}
\def\oIFiveincl{$88.4^{+0.6}_{-0.8}$}
\def\oIFiveRin{$1.28^{+0.02}_{-0.25}$}
\def\oIFivelogxi{$3.8^{+0.2}_{-0.1}$}
\def\oIFiveAfe{$>8$}
\def\oIFivenrel{$5.7^{+1.0}_{-0.1}$}
\def\oIFivenxil{$80^{+10}_{-10}$}
\def\oIFivechi{$86$}
\def\oIFivenu{$67$}
\def\oIFivechired{$1.28$}
\def\oIFiveedgeE{$<4.09$}
\def\oIFiveMaxTau{$0.024^{+0.004}_{-0.003}$}

\def\oISixga{$2.24^{+0.11}_{-0.04}$}
\def\oISixfsc{$0.179^{+0.009}_{-0.009}$}
\def\oISixkTe{$15.8^{+8.5}_{-0.4}$}
\def\oISixTin{$1.049^{+0.007}_{-0.001}$}
\def\oISixndisk{$3.34^{+0.07}_{-0.09}\times10^4$}
\def\oISixq{$0.1$}
\def\oISixincl{$>87$}
\def\oISixRin{$1.28^{+0.03}_{-0.03}$}
\def\oISixlogxi{$3.59^{+0.45}_{-0.04}$}
\def\oISixAfe{$>9.7$}
\def\oISixnrel{$4.08^{+2.60}_{-0.07}$}
\def\oISixnxil{$90^{+280}_{-10}$}
\def\oISixchi{$88$}
\def\oISixnu{$65$}
\def\oISixchired{$1.35$}
\def\oISixedgeE{$4.33^{+0.06}_{-0.06}$}
\def\oISixMaxTau{$0.033^{+0.003}_{-0.003}$}

\def\oISevenga{$>3.35$}
\def\oISevenfsc{$<0.006$}
\def\oISevenkTe{$<24$}
\def\oISevenTin{$1.121^{+0.001}_{-0.001}$}
\def\oISevenndisk{$2.91^{+0.04}_{-0.02}\times10^3$}
\def\oISevenq{$0.1$}
\def\oISevenincl{unconstrained}
\def\oISevenRin{$2.4^{+0.3}_{-0.2}$}
\def\oISevenlogxi{$<0.64$}
\def\oISevenAfe{$5.1^{+1.5}_{-0.2}$}
\def\oISevennrel{$1.17^{+1.02}_{-0.03}$}
\def\oISevennxil{$...$}
\def\oISevenchi{$66$}
\def\oISevennu{$36$}
\def\oISevenchired{$1.83$}
\def\oISevenedgeE{$4.6^{+0.1}_{-0.1}$}
\def\oISevenMaxTau{$0.024^{+0.008}_{-0.008}$}
\def\oISevenEabs{$6.77^{+0.07}_{-0.03}$}
\def\oISevenStrength{$0.3^{+0.3}_{-0.1}$}


\def\oIIOnega{$1.58^{+0.01}_{-0.01}$}
\def\oIIOnefsc{$0.29^{+0.27}_{-0.08}$}
\def\oIIOnekTe{$>200$}
\def\oIIOneTin{$0.022^{+0.003}_{-0.003}$}
\def\oIIOnendisk{$<5\times10^8$}
\def\oIIOneq{$0.1$}
\def\oIIOneincl{unconstrained}
\def\oIIOneRin{$3^{+2}_{-1}$}
\def\oIIOnelogxi{$4.06^{+0.08}_{-0.02}$}
\def\oIIOneAfe{$2.4^{+0.2}_{-0.4}$}
\def\oIIOnenrel{$7^{+2}_{-1}$}
\def\oIIOnenxil{$6.6^{+0.4}_{-1.5}$}
\def\oIIOnechi{$72$}
\def\oIIOnenu{$60$}
\def\oIIOnechired{$0.71$}

\def\oIITwoga{$1.680^{+0.011}_{-0.006}$}
\def\oIITwofsc{$<0.8$}
\def\oIITwokTe{$50^{+90}_{-10}$}
\def\oIITwoTin{$0.49^{+0.02}_{-0.08}$}
\def\oIITwondisk{$<2\times10^5$}
\def\oIITwoq{$0.1$}
\def\oIITwoincl{$<50$}
\def\oIITwoRin{$2.0^{+5.7}_{-0.7}$}
\def\oIITwologxi{$4.38^{+0.04}_{-0.07}$}
\def\oIITwoAfe{$>6$}
\def\oIITwonrel{$0.0050^{+0.0030}_{-0.0004}$}
\def\oIITwonxil{...}
\def\oIITwochi{$57$}
\def\oIITwonu{$61$}
\def\oIITwochired{$0.93$}

\def\oIIThreega{$2.063^{+0.007}_{-0.007}$}
\def\oIIThreefsc{$0.48^{+0.02}_{-0.04}$}
\def\oIIThreekTe{$>220$}
\def\oIIThreeTin{$0.638^{+0.008}_{-0.003}$}
\def\oIIThreendisk{$4.0^{+0.1}_{-0.1}\times10^3$}
\def\oIIThreeq{$0.1$}
\def\oIIThreeincl{$52^{+1}_{-1}$}
\def\oIIThreeRin{$1.9^{+0.7}_{-0.5}$}
\def\oIIThreelogxi{$4.31^{+0.02}_{-0.05}$}
\def\oIIThreeAfe{$5.0^{+0.2}_{-0.2}$}
\def\oIIThreenrel{$6.1^{+0.6}_{-0.2}$}
\def\oIIThreenxil{$3.2^{+0.9}_{-0.9}$}
\def\oIIThreechi{$78$}
\def\oIIThreenu{$60$}
\def\oIIThreechired{$1.31$}

\def\oIIFourga{$2.23^{+0.02}_{-0.03}$}
\def\oIIFourfsc{$0.151^{+0.287}_{-0.003}$}
\def\oIIFourkTe{$>200$}
\def\oIIFourTin{$0.876^{+0.002}_{-0.011}$}
\def\oIIFourndisk{$2^{+1}_{-1}\times10^3$}
\def\oIIFourq{$0.1$}
\def\oIIFourincl{${\rm unconstrained}$}
\def\oIIFourRin{$2.1^{+0.2}_{-0.2}$}
\def\oIIFourlogxi{$>4.4$}
\def\oIIFourAfe{${\rm unconstrained}$}
\def\oIIFournrel{$20^{+8}_{-7}$}
\def\oIIFournxil{$...$}
\def\oIIFourchi{$66$}
\def\oIIFournu{$60$}
\def\oIIFourchired{$1.10$}

\def\oIIFivega{$2.246^{+0.002}_{-0.023}$}
\def\oIIFivefsc{$0.101^{+0.077}_{-0.001}$}
\def\oIIFivekTe{$>200$}
\def\oIIFiveTin{$0.848^{+0.001}_{-0.007}$}
\def\oIIFivendisk{$2.71^{+0.18}_{-0.06}\times10^3$}
\def\oIIFiveq{$0.1$}
\def\oIIFiveincl{$>88$}
\def\oIIFiveRin{$2.9^{+0.7}_{-0.3}$}
\def\oIIFivelogxi{$4.55^{+0.01}_{-0.01}$}
\def\oIIFiveAfe{$>9.8$}
\def\oIIFivenrel{$8.91^{+3.40}_{-0.04}$}
\def\oIIFivenxil{$...$}
\def\oIIFivechi{$42$}
\def\oIIFivenu{$59$}
\def\oIIFivechired{$0.71$}

\begin{deluxetable*}{lccccccc}
\tablecaption{Maximum likelihood estimates of all parameters in spectral fitting of the selected PCA data from outburst 1 of \j1550. \label{tab:params1}}
\tablecolumns{8}
\tablehead{
\colhead{Parameters} & 
\colhead{${\rm HR}=0.91$} & 
\colhead{${\rm HR}=0.82$} & 
\colhead{${\rm HR}=0.67$} & 
\colhead{${\rm HR}=0.52$} & 
\colhead{${\rm HR}=0.38$} &
\colhead{${\rm HR}=0.25$} &
\colhead{${\rm HR}=0.09$} 
}
\startdata
$\Gamma$ & \oIOnega\ & \oITwoga\ & \oIThreega\ & \oIFourga\ & \oIFivega\ & \oISixga\ & \oISevenga\  \\
$F_{\rm sc}$ & \oIOnefsc\ & \oITwofsc & \oIThreefsc\ & \oIFourfsc\ & \oIFivefsc\ & \oISixfsc\ & \oISevenfsc\ \\
$kT_{\rm e}$~[keV] & \oIOnekTe\ & \oITwokTe & \oIThreekTe\ & \oIFourkTe\ & \oIFivekTe\ & \oISixkTe\ & \oISevenkTe\ \\
$T_{\rm in}$~[keV] & \oIOneTin & \oITwoTin\ & \oIThreeTin\ & \oIFourTin\ & \oIFiveTin\ & \oISixTin\ & \oISevenTin\ \\
$N_{\rm disk}$ & \oIOnendisk & \oITwondisk\ & \oIThreendisk\ & \oIFourndisk\ & \oIFivendisk\ & \oISixndisk\ & \oISevenndisk\ \\
$i$~[$^\circ$] & \oIOneincl & \oITwoincl\ & \oIThreeincl\ & \oIFourincl\ & \oIFiveincl\ & \oISixincl\ & \oISevenincl\ \\
$R_{\rm in}~[R_{\rm ISCO}]$ & \oIOneRin & \oITwoRin\ & \oIThreeRin\ & \oIFourRin\ & \oIFiveRin\ & \oISixRin\ & \oISevenRin\ \\
$\log{\xi}$~[${\rm erg~cm^2~s^{-1}}$] & \oIOnelogxi & \oITwologxi\ & \oIThreelogxi\ & \oIFourlogxi\ & \oIFivelogxi\ & \oISixlogxi\ & \oISevenlogxi\ \\
$A_{\rm Fe}$~[Solar] & \oIOneAfe & \oITwoAfe\ & \oIThreeAfe\ & \oIFourAfe\ & \oIFiveAfe\ & \oISixAfe\ & \oISevenAfe\ \\
$N_{\rm rel}~[10^{-3}]$ & \oIOnenrel & \oITwonrel\ & \oIThreenrel\ & \oIFournrel\ & \oIFivenrel\ & \oISixnrel\ & \oISevennrel\ \\
$N_{\rm xil}~[10^{-3}]$ & \oIOnenxil &\oITwonxil\ & \oIThreenxil\ & \oIFournxil\ & \oIFivenxil\ & \oISixnxil\ & \oISevennxil\ \\
$E_{\rm Edge}$~[keV] & \nodata & \nodata & \nodata & \nodata & \oIFiveedgeE\ & \oISixedgeE\ & \oISevenedgeE\ \\
$\tau_{\rm Edge}$ & \nodata & \nodata & \nodata & \nodata & \oIFiveMaxTau\ & \oISixMaxTau\ & \oISevenMaxTau\ \\
$E_{\rm abs}$~[keV] & \nodata & \nodata & \nodata & \nodata & \nodata & \nodata & \oISevenEabs\ \\
${\rm Strength_{abs}}$ & \nodata & \nodata & \nodata & \nodata & \nodata & \nodata & \oISevenStrength\ \\
\hline
$\chi^2$ & \oIOnechi\ & \oITwochi\ & \oIThreechi\ & \oIFourchi\ & \oIFivechi\ & \oISixchi\ & \oISevenchi\  \\
$\nu$ & \oIOnenu\ & \oITwonu\ & \oIThreenu\ & \oIFournu\ & \oIFivenu\ & \oISixnu\ & \oISevennu\ \\
$\chi_{\nu}^2$ & \oIOnechired & \oITwochired & \oIThreechired & \oIFourchired & \oIFivechired & \oISixchired & \oISevenchired \\
\enddata
\tablecomments{$E_{\rm Edge}$ is the xenon L edge energy, $\tau_{\rm Edge}$ is the optical depth of the xenon layer, $E_{\rm abs}$ is the centroid energy of the Gaussian absorption line, representing the ionized disk wind, and ${\rm Strength_{abs}}$ is the strength of that absorption line. The disk normalization is given by $N_{\rm disk} = (R_{\rm in}/D_{10})^2\cos\theta$, where $R_{\rm in}$ is the apparent inner disk in km, $D_{10}$ is the distance to the source in units of 10~kpc, and $\theta$ is the disk inclination. The total $\chi^2$ is shown for each fit, along with the degrees of freedom, $\nu$, and the reduced $\chi^2$, $\chi^2_{\nu}=\chi^2/\nu$. The ionization, $\log\xi$, is given by $L/nR^2$, where $L$ is the ionizing luminosity, $n$ is the gas density, and $R$ is the distance to the ionizing source. All other parameters are as described in the text.}
\end{deluxetable*}

\begin{deluxetable*}{lccccc}
\tablecaption{Maximum likelihood estimates of all parameters in spectral fitting of the selected PCA data from outburst 2 of \j1550. \label{tab:params2}}
\tablecolumns{6}
\tablehead{
\colhead{Parameters} & 
\colhead{${\rm HR}=0.82$} & 
\colhead{${\rm HR}=0.72$} & 
\colhead{${\rm HR}=0.52$} & 
\colhead{${\rm HR}=0.38$} & 
\colhead{${\rm HR}=0.30$} 
}
\startdata
$\Gamma$ & \oIIOnega\ & \oIITwoga\ & \oIIThreega\ & \oIIFourga\ & \oIIFivega\  \\
$F_{\rm sc}$ & \oIIOnefsc\ & \oIITwofsc & \oIIThreefsc\ & \oIIFourfsc\ & \oIIFivefsc\ \\
$kT_{\rm e}$~[keV] & \oIIOnekTe\ & \oIITwokTe & \oIIThreekTe\ & \oIIFourkTe\ & \oIIFivekTe\  \\
$T_{\rm in}$~[keV] & \oIIOneTin & \oIITwoTin\ & \oIIThreeTin\ & \oIIFourTin\ & \oIIFiveTin\  \\
$N_{\rm disk}$ & \oIIOnendisk & \oIITwondisk\ & \oIIThreendisk\ & \oIIFourndisk\ & \oIIFivendisk\ \\
$i$~[$^{\circ}$] & \oIIOneincl & \oIITwoincl\ & \oIIThreeincl\ & \oIIFourincl\ & \oIIFiveincl\  \\
$R_{\rm in}~[R_{\rm ISCO}]$ & \oIIOneRin & \oIITwoRin\ & \oIIThreeRin\ & \oIIFourRin\ & \oIIFiveRin\ \\
$\log{\xi}$~[${\rm erg~cm^2~s^{-1}}$] & \oIIOnelogxi & \oIITwologxi\ & \oIIThreelogxi\ & \oIIFourlogxi\ & \oIIFivelogxi\  \\
$A_{\rm Fe}$~[Solar] & \oIIOneAfe & \oIITwoAfe\ & \oIIThreeAfe\ & \oIIFourAfe\ & \oIIFiveAfe\  \\
$N_{\rm rel}~[10^{-3}]$ & \oIIOnenrel & \oIITwonrel\ & \oIIThreenrel\ & \oIIFournrel\ & \oIIFivenrel\  \\
$N_{\rm xil}~[10^{-3}]$ & \oIIOnenxil &\oIITwonxil\ & \oIIThreenxil\ & \oIIFournxil\ & \oIIFivenxil\  \\
\hline
$\chi^2$ & \oIIOnechi\ & \oIITwochi\ & \oIIThreechi\ & \oIIFourchi\ & \oIIFivechi\  \\
$\nu$ & \oIIOnenu\ & \oIITwonu\ & \oIIThreenu\ & \oIIFournu\ & \oIIFivenu\ \\
$\chi_{\nu}^2$ & \oIIOnechired & \oIITwochired & \oIIThreechired & \oIIFourchired & \oIIFivechired  \\
\hline
\enddata
\end{deluxetable*}


\def\oIIOnega{$1.59^{+0.03}_{-0.02}$}
\def\oIIOnefsc{$0.3^{+0.7}_{-0.2}$}
\def\oIIOnekTe{$80^{+70}_{-30}$}
\def\oIIOneTin{$0.023^{+0.007}_{-0.006}$}
\def\oIIOnendisk{$5^{+500}_{-2}\times10^8$}
\def\oIIOneq{$0.1$}
\def\oIIOneincl{$40^{+10}_{-10}$}
\def\oIIOneRin{$<3$}
\def\oIIOnelogxi{$4.08^{+0.08}_{-0.04}$}
\def\oIIOneAfe{$3.6^{+0.6}_{-0.8}$}
\def\oIIOnenrel{$5^{+2}_{-1}$}
\def\oIIOnenxil{$5^{+2}_{-1}$}
\def\oIIOnechi{$165$}
\def\oIIOnenu{$131$}
\def\oIIOnechired{$1.25$}

\def\oIITwoga{$1.658^{+0.009}_{-0.053}$}
\def\oIITwofsc{$0.5^{+0.3}_{-0.3}$}
\def\oIITwokTe{$70^{+10}_{-10}$}
\def\oIITwoTin{$0.52^{+0.02}_{-0.08}$}
\def\oIITwondisk{$3^{+1}_{-2}\times10^3$}
\def\oIITwoq{$0.1$}
\def\oIITwoincl{$<35$}
\def\oIITwoRin{$1.9^{+2.5}_{-0.5}$}
\def\oIITwologxi{$4.39^{+0.02}_{-0.11}$}
\def\oIITwoAfe{$>7$}
\def\oIITwonrel{$5.6^{+2.2}_{-0.8}$}
\def\oIITwonxil{...}
\def\oIITwochi{$102$}
\def\oIITwonu{$118$}
\def\oIITwochired{$0.87$}

\def\oIIThreega{$2.036^{+0.011}_{-0.002}$}
\def\oIIThreefsc{$0.510^{+0.022}_{-0.006}$}
\def\oIIThreekTe{$>270$}
\def\oIIThreeTin{$0.686^{+0.008}_{-0.006}$}
\def\oIIThreendisk{$2.94^{+0.03}_{-0.08}\times10^3$}
\def\oIIThreeq{$0.1$}
\def\oIIThreeincl{$47^{+3}_{-2}$}
\def\oIIThreeRin{$2.1^{+0.8}_{-0.4}$}
\def\oIIThreelogxi{$4.42^{+0.03}_{-0.05}$}
\def\oIIThreeAfe{$>8$}
\def\oIIThreenrel{$4.6^{+0.2}_{-0.1}$}
\def\oIIThreenxil{$2.2^{+0.9}_{-0.6}$}
\def\oIIThreechi{$150$}
\def\oIIThreenu{$119$}
\def\oIIThreechired{$1.26$}

\def\oIIFourga{$2.245^{+0.002}_{-0.007}$}
\def\oIIFourfsc{$0.349^{+0.012}_{-0.002}$}
\def\oIIFourkTe{$>250$}
\def\oIIFourTin{$0.905^{+0.003}_{-0.017}$}
\def\oIIFourndisk{$2.6^{+0.2}_{-0.1}\times10^3$}
\def\oIIFourq{$0.1$}
\def\oIIFourincl{${\rm unconstrained}$}
\def\oIIFourRin{$1.66^{+0.14}_{-0.06}$}
\def\oIIFourlogxi{$4.15^{+0.2}_{-0.02}$}
\def\oIIFourAfe{$>9.7$}
\def\oIIFournrel{$8.2^{+1.7}_{-0.1}$}
\def\oIIFournxil{$...$}
\def\oIIFourchi{$171$}
\def\oIIFournu{$109$}
\def\oIIFourchired{$1.57$}

\def\oIIFivega{$2.246^{+0.003}_{-0.017}$}
\def\oIIFivefsc{$0.212^{+0.003}_{-0.001}$}
\def\oIIFivekTe{$>280$}
\def\oIIFiveTin{$0.868^{+0.002}_{-0.009}$}
\def\oIIFivendisk{$2.877^{+0.008}_{-0.094}\times10^3$}
\def\oIIFiveq{$0.1$}
\def\oIIFiveincl{${\rm unconstrained}$}
\def\oIIFiveRin{$1.9^{+0.1}_{-0.1}$}
\def\oIIFivelogxi{$3.82^{+0.02}_{-0.02}$}
\def\oIIFiveAfe{$>9.8$}
\def\oIIFivenrel{$4.10^{+1}_{-0.05}$}
\def\oIIFivenxil{$...$}
\def\oIIFivechi{$228$}
\def\oIIFivenu{$110$}
\def\oIIFivechired{$2.07$}

\begin{deluxetable*}{lccccc}
\tablecaption{Maximum likelihood estimates of all parameters in spectral fitting of the selected PCA and HEXTE data from outburst 2 of \j1550. \label{tab:params2_2}}
\tablecolumns{6}
\tablehead{
\colhead{Parameters} & 
\colhead{${\rm HR}=0.82$} & 
\colhead{${\rm HR}=0.72$} & 
\colhead{${\rm HR}=0.52$} & 
\colhead{${\rm HR}=0.38$} & 
\colhead{${\rm HR}=0.30$} 
}
\startdata
$\Gamma$ & \oIIOnega\ & \oIITwoga\ & \oIIThreega\ & \oIIFourga\ & \oIIFivega\  \\
$F_{\rm sc}$ & \oIIOnefsc\ & \oIITwofsc & \oIIThreefsc\ & \oIIFourfsc\ & \oIIFivefsc\ \\
$kT_{\rm e}$~[keV] & \oIIOnekTe\ & \oIITwokTe & \oIIThreekTe\ & \oIIFourkTe\ & \oIIFivekTe\  \\
$T_{\rm in}$~[keV] & \oIIOneTin & \oIITwoTin\ & \oIIThreeTin\ & \oIIFourTin\ & \oIIFiveTin\  \\
$N_{\rm disk}$ & \oIIOnendisk & \oIITwondisk\ & \oIIThreendisk\ & \oIIFourndisk\ & \oIIFivendisk\ \\
$i$~[$^{\circ}$] & \oIIOneincl & \oIITwoincl\ & \oIIThreeincl\ & \oIIFourincl\ & \oIIFiveincl\  \\
$R_{\rm in}~[R_{\rm ISCO}]$ & \oIIOneRin & \oIITwoRin\ & \oIIThreeRin\ & \oIIFourRin\ & \oIIFiveRin\ \\
$\log{\xi}$~[${\rm erg~cm^2~s^{-1}}$] & \oIIOnelogxi & \oIITwologxi\ & \oIIThreelogxi\ & \oIIFourlogxi\ & \oIIFivelogxi\  \\
$A_{\rm Fe}$~[Solar] & \oIIOneAfe & \oIITwoAfe\ & \oIIThreeAfe\ & \oIIFourAfe\ & \oIIFiveAfe\  \\
$N_{\rm rel}~[10^{-3}]$ & \oIIOnenrel & \oIITwonrel\ & \oIIThreenrel\ & \oIIFournrel\ & \oIIFivenrel\  \\
$N_{\rm xil}~[10^{-3}]$ & \oIIOnenxil &\oIITwonxil\ & \oIIThreenxil\ & \oIIFournxil\ & \oIIFivenxil\  \\
\hline
$\chi^2$ & \oIIOnechi\ & \oIITwochi\ & \oIIThreechi\ & \oIIFourchi\ & \oIIFivechi\  \\
$\nu$ & \oIIOnenu\ & \oIITwonu\ & \oIIThreenu\ & \oIIFournu\ & \oIIFivenu\ \\
$\chi_{\nu}^2$ & \oIIOnechired & \oIITwochired & \oIIThreechired & \oIIFourchired & \oIIFivechired  \\
\hline
\enddata
\end{deluxetable*}

\section{Discussion}
\label{sec:discussion}

The results of our reflection modeling of \j1550\ can be summarized as follows. The reflection features of \j1550\ appear to display stronger blueward emission as the source transitions to the soft state. In the very soft branch during outburst 1 in 1998/99, we see tentative evidence for an absorption line at $\sim6.9$~keV, possibly associated with \ion{Fe}{26} in an ionized disk wind. Throughout outburst 1, the accretion disk is constrained from the reflection component to be within a few times the ISCO (up to $8\times~R_{\rm ISCO}$ during outburst 2), and broadly consistent with being within a factor of $1$--$2~R_{\rm ISCO}$. The coronal temperature is higher during outburst 2 in 2000, consistent with less IC cooling, and expected given the comparatively lower X-ray luminosity with respect to outburst 1. During hard states, the disk inclination constraints are low, and roughly consistent with the value determined by C19 in modeling of the bright hard-intermediate state. As the source progresses to softer states, we see evidence for much higher disk inclinations (typical values of $>85^{\circ}$). However, reflection models with purely IC irradiation spectra (e.g., {\tt relxillCp}) do not provide good quality fits to the data in the softer states. In the following Sections (\ref{sec:evolution}, \ref{sec:return}, \ref{sec:inclination}) we discuss a comparison of our overall results with previous global studies of the outbursts of \j1550\ as well as other BHBs, the disk inclination discrepancy, and explore the complications which arise when fitting a reflected IC component to the soft state data.

\subsection{The evolution of \j1550}
\label{sec:evolution}

\cite{Sobczak2000} provided a comprehensive study of the spectral variation of the first detected outburst of \j1550\ in 1998/99, using \rxte\ (PCA and HEXTE) spectral data. Through a more empirical treatment of disk+power law emission (with the Fe K emission and smeared absorption edges included), they determined a range for the power law cutoff energy of $\sim20\mbox{--}50$~keV during the initial rise of the hard state. We find, consistently with them, a coronal temperature range of $\sim10\mbox{--}35$~keV during this same phase, corresponding to roughly $20$--$105$~keV in the cutoff energy (assuming $E_{\rm cut}\sim2\mbox{--}3~kT_{\rm e}$, an approximate range for the cutoff due to dispersion, geometrical, and relativistic effects, e.g., \citealt{Petrucci2001}).


\begin{figure*}
\centering
\includegraphics[width=0.7\linewidth]{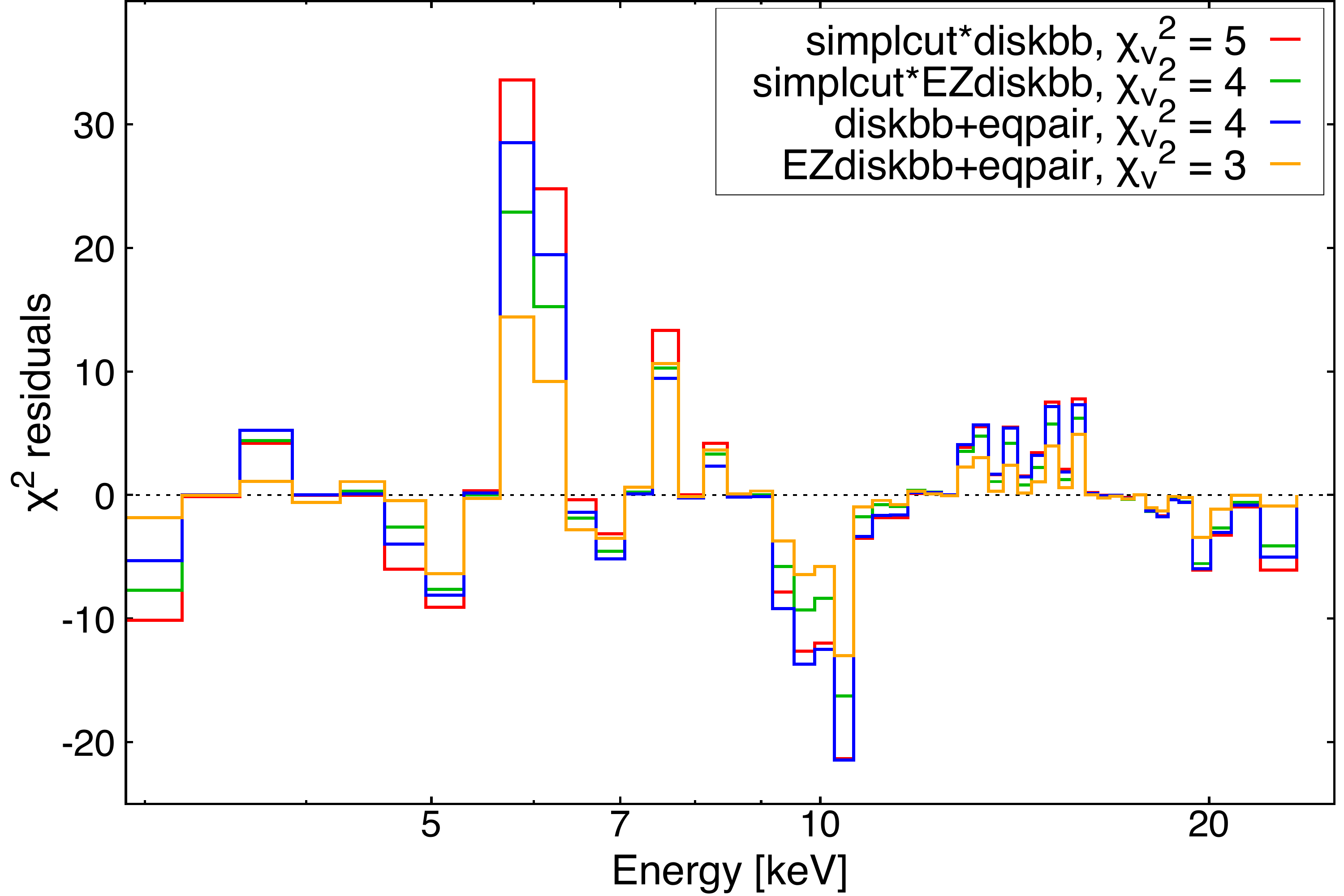}
\caption{$\chi^2$ residuals of different continuum model fits to the 40401-01-27-00 PCU~2 spectra. The key shows each continuum component where the total model is {\tt\justify crabcorr * TBabs * (CONTINUUM) * edge * gabs}. }
\label{fig:diff_models}
\end{figure*}

\cite{Sobczak2000} also found that the inner disk radius decreased sharply at the time of the 6.8~Crab flare during outburst 1, by over an order of magnitude. They attributed this to the inaccuracy of the blackbody disk model applied to the data. We chose not to include the flare data in our small sample, which was based upon the apparent difference in the nature of the power-law-like emission during the flare---X-ray emission could be originating in a jet component (e.g., \citealt{mnw05}). However, we do see tentative evidence for curious variations in the disk inner radius around the time of the flare. Table~\ref{tab:params1} and Figure~\ref{fig:pars} show that at a hardness ratio of 0.52, the disk radius appears to have increased with respect to the disk at HR = 0.67 (from $<3~R_{\rm ISCO}$ to $3.7^{+0.5}_{-0.4}~R_{\rm ISCO}$), and the disk radius then decreases at HR = 0.38 ($1.28^{+0.02}_{-0.25}~R_{\rm ISCO}$). The observation at HR = 0.52 was made pre-flare (MJD~51071.2), which occurred between MJDs $\sim51072$ and $51080$. The close proximity to the bright flare could perhaps indicate that the flare was associated with a slight recession of the disk. The observation at ${\rm HR}=0.38$ was made approximately 30 days after the flare (MJD~51108.1), and whilst a statistical distinction cannot be made between the disk radius at this time, and in the harder states, it is nonetheless the case that the disk appears closer to the ISCO after the flare, than moments before it. Given that we do not perform a detailed analysis of all the observations surrounding, and during the flare, we cannot make stronger independent statements regarding the rapid movement of the inner disk. Our overall characterization of the inner disk radius throughout outburst 1 of \j1550 agrees well with the results found by \cite{Sobczak2000}, always remaining within tens of $r_{\rm g}$.

The second outburst of \j1550\ as tracked by \rxte\ was studied in a similar way by \cite{Rodriguez2003}. They generally find lower coronal cutoff energies than those predicted by our constraints on $kT_{\rm e}$. However, \cite{Rodriguez2003} do not give constraints on the soft-to-intermediate states. In addition, we have reduced systematics using the {\tt pcacorr} tool, and adopted more complex reflection models to achieve more physical constraints from the data. Since we have also checked our constraints when including HEXTE spectra for all the observations in our sample of outburst 2, we suggest the differences between the results of \cite{Rodriguez2003} and our own are mostly due to modeling distinctions.  

In the broader context, our detailed analysis of the evolution of \j1550\ and its reflection and coronal properties agrees very well with previous examples of this type of analysis on GX~339$-$4. For example, \cite{Garcia2015} found that the inner disk radius is already within a few times the ISCO during the bright hard state of GX~339$-$4: we find the same for \j1550. In addition, they showed the trend of decreasing coronal temperature as the source rises in its hard state. Since \j1550\ has not been tracked from the low through to the high hard state, we do not have a very direct comparison with the results of \cite{Garcia2015} with regards to the coronal temperature. However, we do nonetheless see the continuing evolution of the coronal temperature, from high to low, and most importantly, we have shown that reflection modeling of this kind demonstrates the clear luminosity dependence: a more luminous corona is a cooler corona. 

\cite{Sridhar2020} recently explored the evolution of GX~339$-$4 in transition from the hard to the soft state, using a similar type of detailed analysis of its reflection properties. They showed that the inner disk radius remains constant during the transition, having approached the ISCO during the bright hard state. They also showed that the disk is only mildly truncated (within $\sim10~r_{\rm g}$) in all their selected observations. We find results for \j1550\ which are consistent with this, the only key difference being that we do see some evolution in the reflection spectrum of \j1550\ as it transitions into the much softer states (and it is of note that \j1550\ reaches PCA count rates several factors higher than GX~339$-$4 during its softest spectral state), and this is qualitatively shown in Figure~\ref{fig:line}: the Fe K emission appears to show more prominent blueward emission as the spectrum becomes softer. As we discuss in the following Sections, we attribute this evolution either to inclination changes in the inner disk (i.e., a warp), or a signal of the need to evolve our treatment of the dominant irradiative spectrum in the soft state, or indeed both. 

In Section~\ref{sec:refl_results} we showed that the inner disk inclination, as constrained by reflection spectral modeling of the PCA data, appears to increase sharply as the source transitions to the soft state. Whilst in some cases the inclination is poorly constrained (see Figure~\ref{fig:pars} and Tables~\ref{tab:params1} and \ref{tab:params2}), in both outbursts 1 and 2 there is a clear evolution from relatively moderate inclination ($\sim30^{\circ}\mbox{--}50^{\circ}$) to almost maximal inclinations ($90^{\circ}$). Ignoring for now the ubiquitous discrepancy with the binary orbit inclination ($\sim75^{\circ}$; \citealt{Orosz2011}), we suggest that the sudden change in inclination we are deriving is an artifact of the models being applied. The difficulty we have fitting the coronal reflection model {\tt relxillCp} to the very soft state data (see Table~\ref{tab:params1}) is further evidence that the assumed irradiation spectrum for reflection is unrealistic. We find it likely that the evolving reflection spectrum and disk inclination is caused by a shift in the dominant illuminating spectral component. We discuss this in more detail in the following Section.

\subsection{Returning disk radiation}
\label{sec:return}

Before testing alternative reflection models in the soft state, we first show the requirement for reflection in the X-ray spectrum. As a clarification of the requirement of Fe K emission in the soft state, we fit observation 40401-01-27-00 (the softest observation in our sample for outburst 1, ${\rm HR}=0.09$) with a selection of different disk and coronal emission components. This is shown in Figure~\ref{fig:diff_models}. We used the disk blackbody model variant {\tt EZdiskbb} \citep{Zimmerman2005}, which differs from {\tt diskbb} in its boundary conditions, assuming a zero torque at the inner edge of the accretion disk. We also tried an alternative coronal IC emission model, {\tt eqpair} \citep{Coppi2000}. The {\tt eqpair} model is somewhat different since it includes a prescription for a hybrid distribution of electrons, with some fraction of energy going into thermal and non-thermal distributions. The turbulent nature of accretion makes the idea of a purely thermal distribution of high-energy particles unlikely, thus {\tt eqpair} is a more physically consistent treatment of the microphysics of accreting plasmas. We do not test the {\tt nthComp} model here since we already addressed the contrasts with {\tt simplcut} in Section~\ref{sec:refl_results}, showing that there is no difference in the IC continuum, but only in the ability to constrain the disk component (see Figure~\ref{fig:simpl_v_nthComp}). 

\begin{figure*}
\centering
\includegraphics[width=0.45\linewidth]{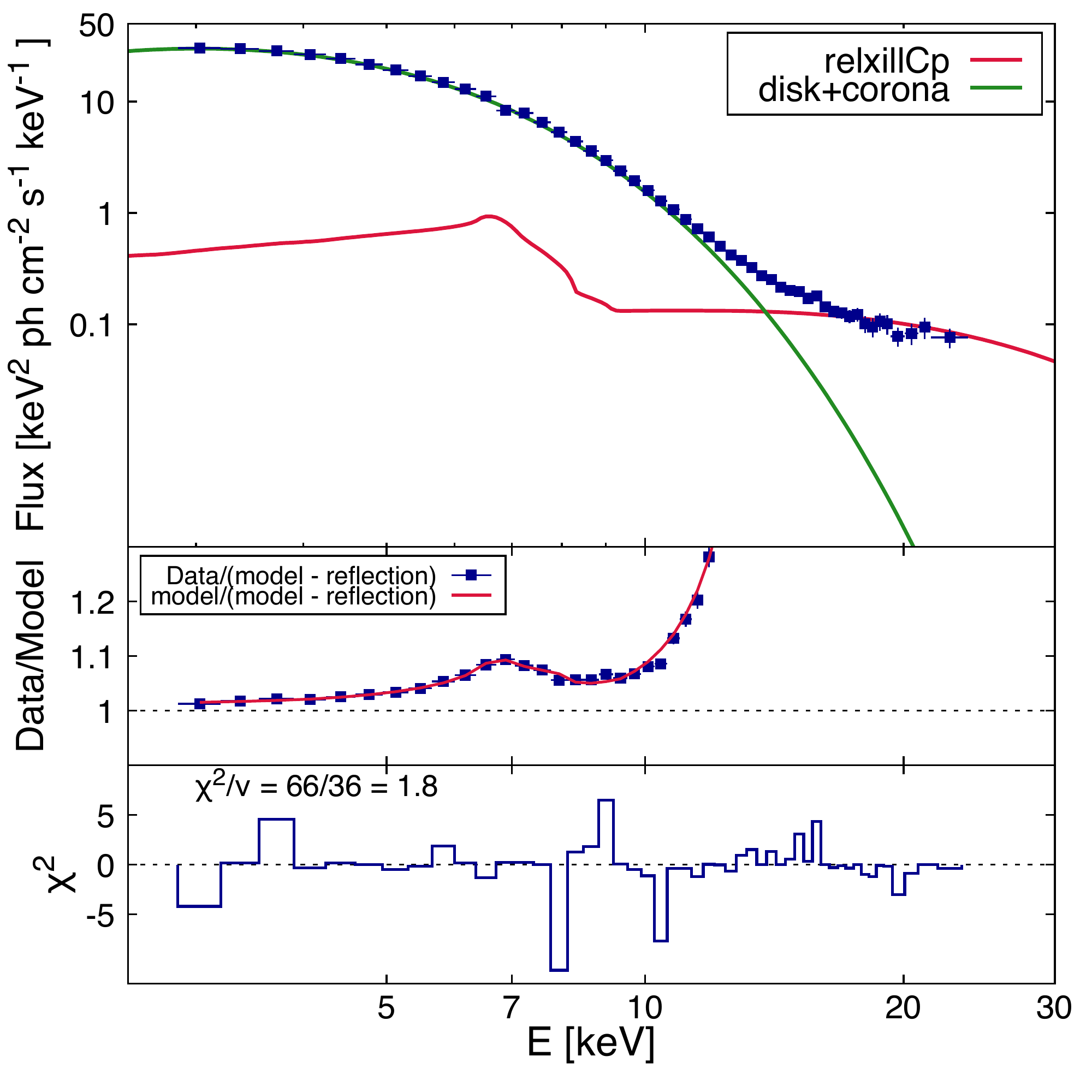}
\includegraphics[width=0.45\linewidth]{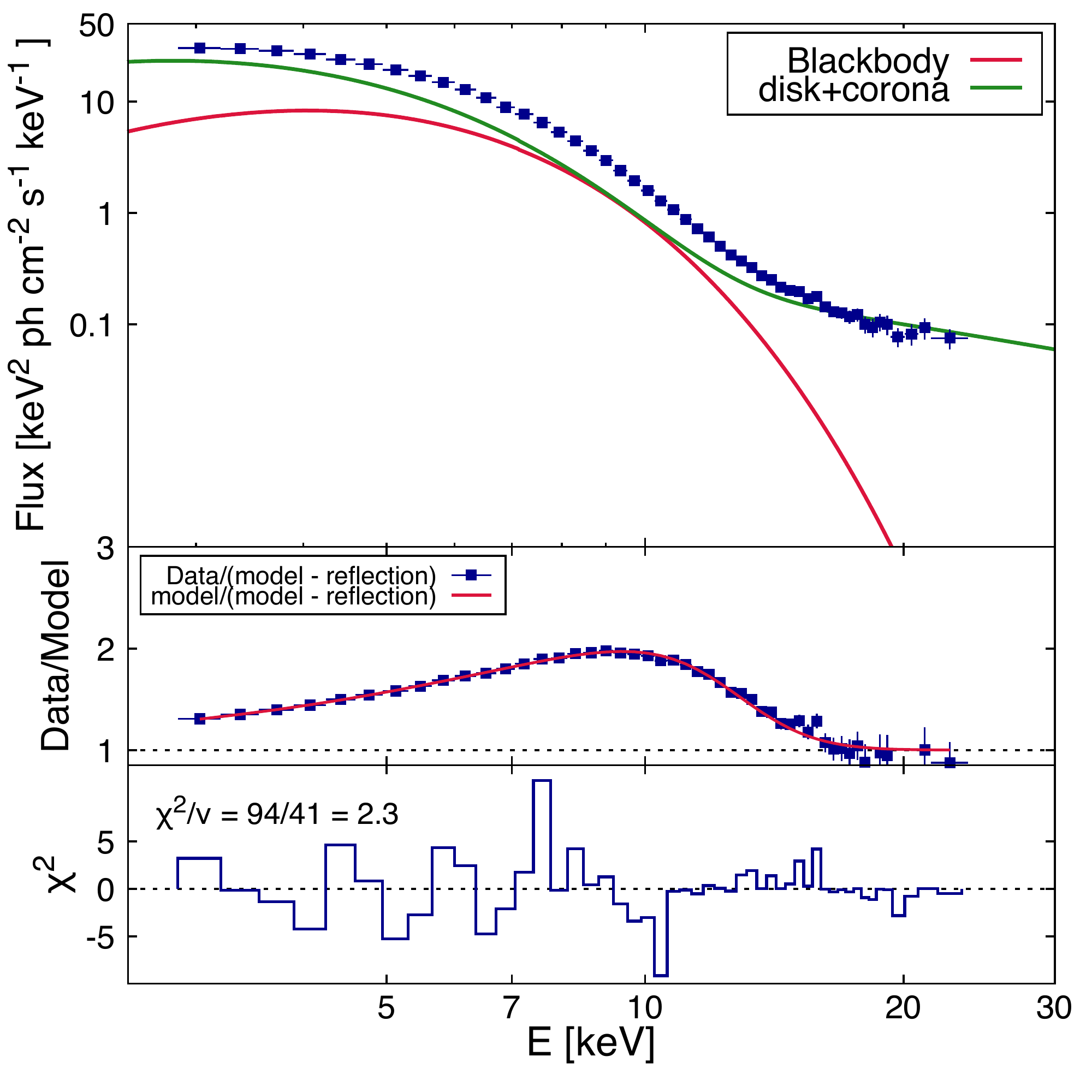}
\includegraphics[width=0.45\linewidth]{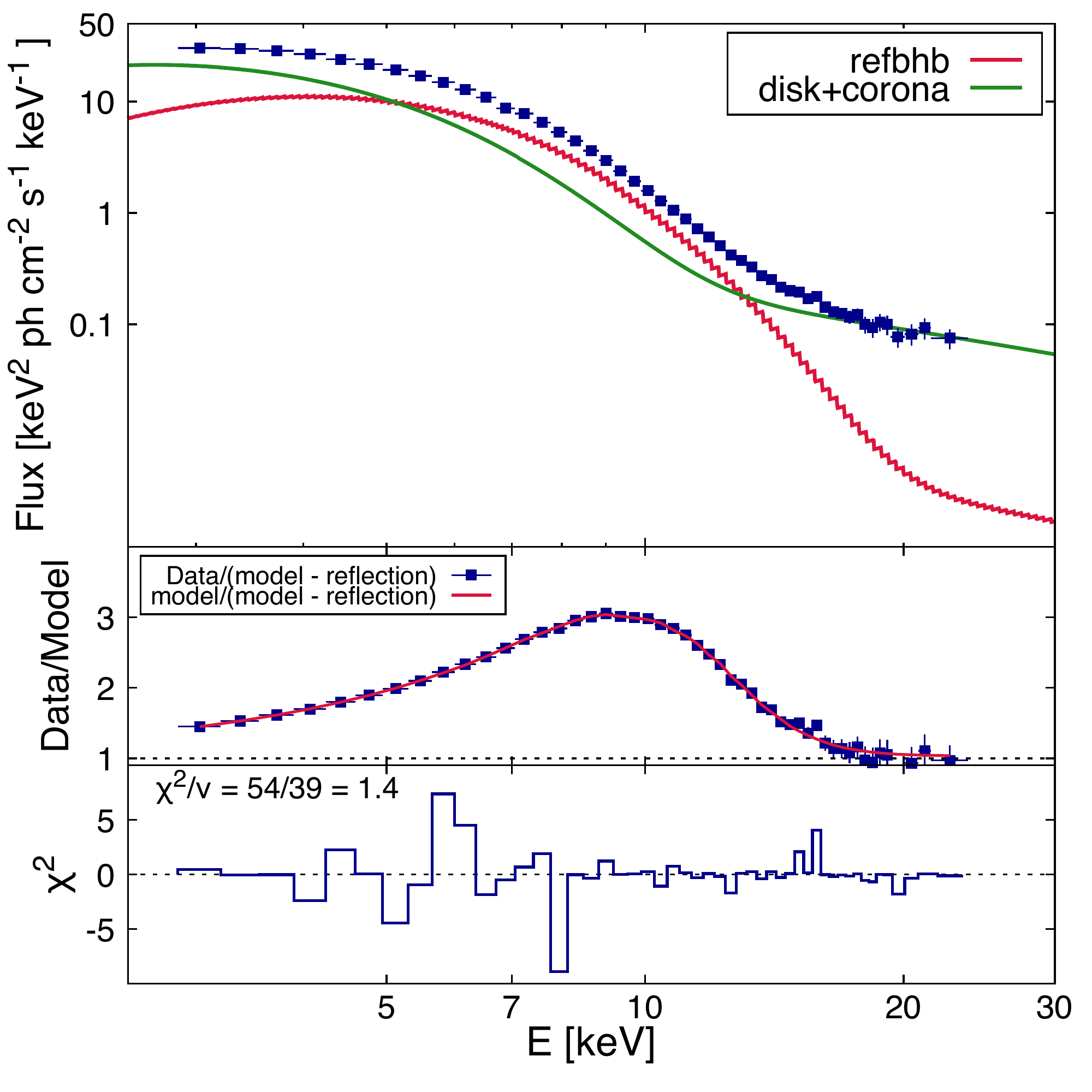}
\includegraphics[width=0.45\linewidth]{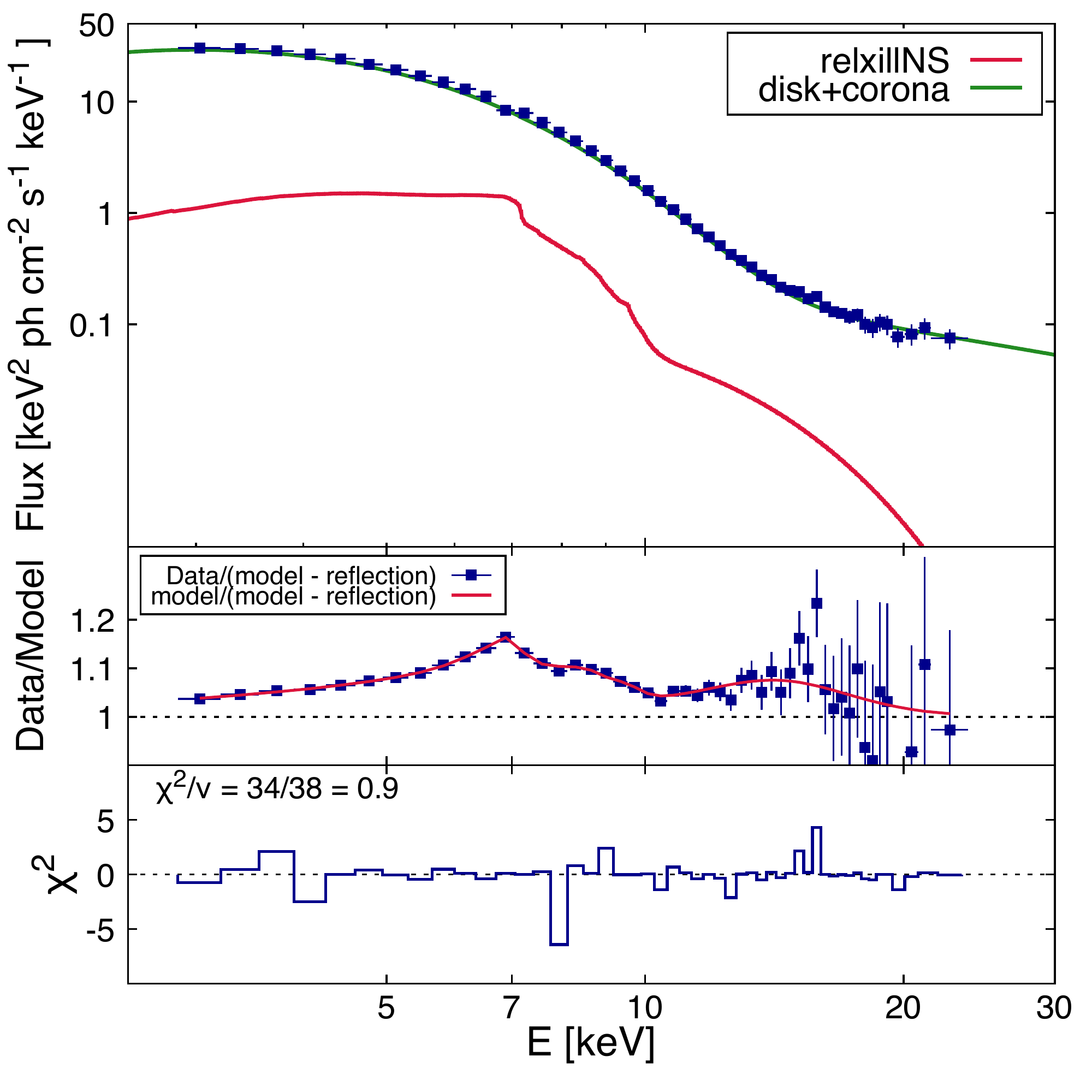}
\caption{A comparison of the models {\tt relxillCp}, {\tt bbody}, {\tt refbhb} and {\tt relxillNS} applied to spectral observation 40401-01-27-00, the softest dataset in our selection at HR = 0.09. The reflection spectrum is more naturally produced as a sub-dominant component resulting from illumination of a blackbody spectrum, as opposed to a harder component with a prominent Compton hump. The reflection component contributes $\sim5$\% of the total flux in the {\tt relxillNS} fit (bottom right panel), and this model is the only one which satisfactorily fits the residuals beyond the disk + corona continuum. }
\label{fig:relxill_comp}
\end{figure*}

We tested the ability of these different continuum models to fit the PCU~2 spectra of observation 40401-01-27-00, in order to check for whether there are significant changes in the resulting residuals. As shown in Figure~\ref{fig:diff_models}, although the more complex disk+corona continuum model of {\tt EZdiskbb + eqpair} leads to some reduction in residual features, in particular the Fe line residuals at low energies ($\sim6$~keV), and the Fe edge in the $10$~keV region, the line and edge features still remain. Thus, reflection is required to fit the soft state spectrum.

We therefore tried an alternative model which adopts a softer, thermal continuum as its irradiating spectrum, {\tt relxillNS} (Garc\'ia et al., in prep). The {\tt relxillNS} model is a flavor of the {\tt relxill} suite of models, developed to prescribe reflection from accretion disks around neutron stars. The irradiating spectrum is a single-temperature blackbody. We expect that representing the irradiating continuum as a single temperature blackbody, as opposed to a multitemperature disk blackbody spectrum, will be sufficient, given that only photons in the very inner regions of the disk will experience light bending effects. The full model is very similar to that shown in the results of Section~\ref{sec:refl_results}, except we replace the {\tt relxillCp} model with {\tt relxillNS}: {\justify\tt crabcorr * TBabs * (simplcut$\otimes$diskbb + relxillNS) * edge * gabs}. Since now the reflection continuum does not relate to the IC spectrum, we fix the coronal temperature to $kT_{\rm e} = 300$~keV. This is based upon initial comparisons of this model with the data, which reveal that the cutoff is not constrained once the Comptonized disk component of the model fits the high-energy tail. In addition, we fix the inner disk radius to $R_{\rm in} = R_{\rm ISCO}$. Again this is a result of initial comparison with the data, revealing that the inner radius is consistent with being at the ISCO, as well as the assumption that the disk is likely not truncated in this very soft state. Therefore, whereas the irradiating continuum of {\tt relxillCp} depends explicitly on the $\Gamma$ and $kT_{\rm e}$ parameters of {\tt simplcut}, the irradiating continuum of {\tt relxillNS} instead depends on the inner disk temperature ($T_{\rm in}$) of {\tt diskbb}, which sets the temperature of the reflected blackbody component.

In order to properly test the robustness of the {\tt relxillNS} model to fit the very soft state \j1550\ data, we directly compared it with other models applied to the data. We began by comparing fits of four different representative reflection/continuum components to model the residuals in the Fe K emission region. We tried four interchangeable components: {\tt relxillCp}, {\tt relxillNS}, {\tt refbhb}, and {\tt bbody}. The {\tt refbhb} model \citep{Ross2007} is a reflection model which includes the underlying emission from the accretion disk. Thus the output spectrum of {\tt refbhb} includes blackbody, and reflected power law components. The {\tt bbody} model is a single-temperature blackbody component. We include this as an alternative to test whether or not explicit reflection features are actually required by the data, i.e., Fe K emission and the Fe smeared edge feature. In addition, since {\tt relxillNS} adopts a single-temperature blackbody as its irradiating continuum, {\tt bbody} represents the pre-reflection continuum of {\tt relxillNS}, and is useful as a direct comparison. The results are shown in Figure~\ref{fig:relxill_comp}.

 It is clear that the only model capable of reproducing the subtleties of the Fe K region, whilst also producing a good overall fit to the continuum, is {\tt relxillNS}. As already discussed, {\tt relxillCp} struggles to capture the line and edge features whilst maintaining an appropriate fit to the broader continuum. This is due to the softness of the spectrum. Since the disk is being irradiated with a coronal IC continuum, any successful fit of {\tt relxillCp} to the reflection features in the data naturally produces a strong Compton hump above 10~keV, and {\tt relxillCp} overfits the continuum at high energies. In contrast, whilst {\tt refbhb} is a much softer reflection model due to the underlying disk emission, it fails to model out the Fe K residuals. This is because the Fe line emission is inherently weaker since the high disk temperature results in an overionized atmosphere. Thus, {\tt refbhb} is effectively only fitting the continuum. The {\tt bbody} model therefore behaves very similarly to {\tt refbhb}, as expected. The {\tt relxillNS} model fits very well to the Fe K region, and since the irradiating continuum is a blackbody, the lack of high-energy irradiating X-rays results in a subdominant Compton hump. The best fit parameters and their uncertainties for the fit to 40401-01-27-00 with {\tt relxillNS} is shown in Table~\ref{tab:soft_state}. The model fits very well to the data, and we do not require any truncation on the disk, with $R_{\rm in}=R_{\rm ISCO}$. The disk inclination is low, $i=37 \pm 4$~degrees, similar to values attained during our fits to hard state data (see Table~\ref{tab:params1}), and matching exactly with the value obtained by C19. We also attempted to model the PCA observation 40401-01-50-00 (${\rm HR}=0.25$) with the {\tt relxillNS} model substituted for {\tt relxillCp}, and struggled to fit the data. It is likely that already at those hardness ratios the spectrum is too IC dominated for the reflection model to be simplified to a blackbody shape. Thus a future hybrid model may be an interesting test to perform in future work.



\def\softga{$3.0^{+0.4}_{-0.4}$}
\def\softfsc{$0.018^{+0.014}_{-0.007}$}
\def\softkTe{$300$\tablenotemark{a}}
\def\softTin{$1.105^{+0.005}_{-0.006}$}
\def\softndisk{$3.03^{+0.07}_{-0.07}\times10^3$}
\def\softq{$0.1$}
\def\softincl{$37^{+4}_{-4}$}
\def\softRin{$1$\tablenotemark{a}}
\def\softlogxi{$2.7^{+0.3}_{-0.2}$}
\def\softAfe{$>5$}
\def\softnrel{$5.1^{+2.7}_{-0.9}$}
\def\softnxil{$...$}
\def\softedgeE{$4.58^{+0.09}_{-0.11}$}
\def\softmaxtau{$0.031^{+0.007}_{-0.007}$}
\def\softlineE{$6.74^{+0.08}_{-0.08}$}
\def\softStrength{$0.5^{+3.0}_{-0.3}$}
\def\softchi{$34$}
\def\softnu{$38$}
\def\softchired{$0.9$}

\begin{deluxetable}{lr}
\tablecaption{Maximum likelihood estimates of all parameters in spectral fitting of observation 40401-01-27-00. \label{tab:soft_state}}
\tablehead{
\colhead{Parameters} & MLEs with 90\% uncertainties\\ 
}

\startdata
$\Gamma$ & \softga\   \\
$F_{\rm sc}$ & \softfsc\ \\
$kT_{\rm e}$~[keV] & \softkTe\   \\
$T_{\rm in}$~[keV] & \softTin   \\
$N_{\rm disk}$ & \softndisk \\
$i$~[$^{\circ}$] & \softincl  \\
$R_{\rm in}~[R_{\rm ISCO}]$ & \softRin  \\
$\log{\xi}$~[$L/nR^2$] & \softlogxi    \\
$A_{\rm Fe}$~[Solar] & \softAfe   \\
$N_{\rm rel}~[10^{-3}]$ & \softnrel   \\
$E_{\rm Edge}$~ [keV] & \softedgeE \\
$\tau_{\rm Edge}$ & \softmaxtau \\
$E_{\rm abs}$~[keV] & \softlineE \\
${\rm Strength_{abs}}$ & \softStrength \\
\hline
$F_{\rm Total}$~[${\rm erg}~{\rm cm}^{-2}~{\rm s}^{-1}$] & $6.62\times10^{-8}$ \\
$F_{\rm Disk+Corona}$~[${\rm erg}~{\rm cm}^{-2}~{\rm s}^{-1}$] & $6.27\times10^{-8}$ \\
$F_{\rm Refl}$~[${\rm erg}~{\rm cm}^{-2}~{\rm s}^{-1}$] & $0.35\times10^{-8}$ \\
$F_{\rm Refl}/F_{\rm Total}$ & $5.2\%$ \\
\hline
$\chi^2$ & \softchi  \\
$\nu$ & \softnu  \\
$\chi_{\nu}^2$ & \softchired   \\
\hline
\enddata
\tablenotetext{a}{Frozen parameter.}
\tablecomments{The model fit to the soft state spectrum is {\tt\justify crabcorr * TBabs * (simplcut$\otimes$diskbb + relxillNS) * edge * gabs}. The edge accounts for xenon in the PCU~2 layers. The {\tt gabs} component represents an absorption line in an ionized disk wind, which we allow to run free between 6--7~keV.}
\end{deluxetable}

\begin{figure}[h!]
\vspace{-50pt}
\includegraphics[width=\linewidth, trim=100 50 100 150, clip]{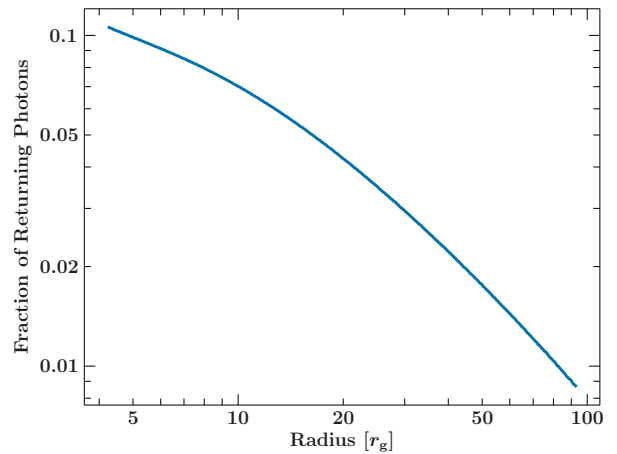}
\vspace{-120pt}
\caption{The fraction of blackbody photons returning from the accretion disk as a function of disk radius. The total fraction of photons which return to the disk is $\sim5.4\%$, assuming a \cite{SS1973} disk.  }
\label{fig:return}
\end{figure}

In order to check the physics of our result that there may be reflection of disk emission in the soft state,
we calculate the fraction of photons we should expect to return to the disk. Employing
the code by \cite{Yang2013}, we perform General relativistic ray
tracing simulations to predict the amount of photons returning to the
disk. 

Using a standard prescription for sub-critical accretion from the disk (in the Newtonian approximation) 

\begin{equation}
Q^+(R) = \frac{3}{8\pi} \frac{G M \dot{M} }{R^3} \left(1 - \sqrt{\frac{R_\mathrm{in}}{R}} \right)
\end{equation}

\noindent \citep{SS1973,Poutanen2007}, we can estimate the total
fraction of photons returning to the disk as $\sim5.4\%$, assuming a BH spin $a^*=0.5$. Figure~\ref{fig:return} shows the fraction of disk photons that return to
the disk as a function of disk radius. We assume that the disk
reaches all the way to the ISCO, in accordance with the fit results shown in Table~\ref{tab:soft_state}. This fraction of $\sim5.4\%$
is well matched with the fractional flux required by the {\tt relxillNS}
reflection component in order to fit the soft state spectrum of \j1550 ($\sim5.2\%$). Therefore we consider the reflection by self-irradiation a
valid explanation of the observed spectrum. However, we stress that our implementation of the {\tt relxillNS} model is not a final, developed implementation of returning disk radiation, and is not intended as such, it is instead an approximate first representation. A detailed
characterization of the reflection from returning radiation, requires
a more complex treatment, including exhaustive spectral energy shifts
from each part of the disk, which is beyond the scope of this paper. The evidence in \cite{Steiner2016} for stronger Fe lines in soft states relative to hard states with similar power-law flux is consistent with our prediction and expectation.  Future developments of the returning radiation model may enable direct testing of this scaling against the full RXTE BHB database.


\subsection{Disk inclination discrepancy}
\label{sec:inclination}

C19 showed, through reflection modeling of a simultaneous \asca\ and \rxte\ observation of \j1550\ during the hard-intermediate state, that the inclination of the inner accretion disk appears to be much lower than the binary inclination ($\sim39^{\circ}$ as opposed to $\sim75^{\circ}$). Our broader analysis of \j1550\ here has shown that this was not an anomaly. Reflection modeling of the hard state of \j1550\ with the model {\tt relxillCp} shows that the disk inclination is low in the hard and hard-intermediate states (HR $\gtrsim0.5$). In addition, once adopting a more appropriate irradiating continuum in the soft state (see Section~\ref{sec:return}), we obtain a lower inclination ($\sim37^{\circ}$), consistent with the hard state modeling, and the results of C19. 

There are several possible reasons for this apparent discrepancy, some of which were addressed by C19: (i) we are detecting a very significant warp in the accretion disk of \j1550; (ii) the inner disk structure has a vertical structure which could be giving rise to obscuration of the blueward line emission, thus leading to an inferred inclination much lower than the true value; (iii) the irradiating source is actually an outflowing, relativistic jet, thus altering the shape of the irradiating flux with respect to a static source; (iv) the disk density is much higher than the assumed value in our modeling ($n_{\rm e}=10^{15}~{\rm cm^{-3}}$). 

The discovery of an apparent ionized disk wind in the soft state secures the fact that at least the outer portion of the accretion disk must be at high inclination. This is because disk winds should not, and indeed ubiquitously are not, detectable in low-inclination BHBs \citep{Ponti2012}. This supports the idea that our reflection modeling constraints are only tracking the inclination of the inner disk. Therefore, we must either be detecting a warped inner disk, or effects not included in our modeling are acting to skew our estimates to low inclination. 

%
%
%
\section{Conclusions}
\label{sec:conclusion}

We have presented results of reflection modeling of a sample of \rxte-PCA data from observations of the first two outbursts of \j1550\, covering the hard-to-soft spectral states. We find several key results. The global evolution of \j1550\ is consistent with the picture that the inner disk radius, assuming a BH spin $a_{\star}=0.5$, is only slightly truncated (within a few times the ISCO) during the bright hard state, moving inwards to the ISCO in transition to the soft state (e.g., \citealt{Garcia2015,Sridhar2020}). 


During the very soft branch of outburst 1, \j1550\ shows possible evidence for an ionized disk wind via a ubiquitous absorption feature at $\sim6.9$~keV. This feature has not been detected in \j1550 X-ray observations during this soft branch before (e.g., \citealt{Sobczak2000}), however, it is common to detect such wind features in soft-state BHBs (e.g., \citealt{Lee2002, Miller2006a, Ponti2012}). 

We have confirmed that the low disk inclination obtained by C19 in modeling of the hard-intermediate state of \j1550\ was not an anomalous result: in the hard state we typically measure low inclinations coinciding with the value found by C19. In the soft state, assuming coronal IC is the dominant irradiating component, reflection modeling yields unreasonably high disk inclinations, close to $90^{\circ}$. We conclude that the assumed irradiating continuum, i.e. coronal IC emission, is inadequate for reflection models of the soft state. However, since we still obtain inclination estimates which are mismatched with the binary inclination of $75^{\circ}$, \j1550\ may be an example of a BHB with a warped disk. Alternatively, as presented by C19, and explored theoretically by \cite{Taylor2018}, the vertical structure of the inner disk may be obscuring blueward line emission, leading to lower inferred disk inclinations in reflection modeling. Strong density effects in the disk, as well as more complex irradiating source geometries, could also contribute to the inclination estimates. These are all phenomena that are the focus of future work in the field of relativistic reflection modeling. 

The most remarkable result of our analysis is the first apparent detection of self-irradiating disk reflection. We find that during the very soft states, when the disk blackbody emission dominates the X-ray spectrum, the reflection spectrum is likely being produced by self-irradiating blackbody disk photons, $\sim5\%$ of which we expect to return to the inner disk. We showed that initial calculations of the proportion of photons we expect to return to the inner disk regions, assuming a BH spin of $a_{\rm \star}=0.5$, are comparable with the fraction of overall flux in the reflection component found from our modeling. As such, we suggest that as BHBs transition from the hard to soft states, models should necessarily include disk emission as an irradiative component for reflection. The development of such a model for a full relativistic, self-consistent treatment of returning disk radiation will be the subject of a future paper. This future model will be a more improved version of {\tt relxillNS}, in which the appropriate multitemperature disk blackbody spectrum is assumed, and the emissivity profile of the returning radiation is self-consistently calculated via GR ray tracing simulations for a given BH spin and inner disk radius (in a similar way to the simulations used to generate the {\tt relxill} suite of models). We will be able to apply this more self-consistent model to data across the hard and soft states of BHBs, thus allowing us to make more direct comparisons of the reflection properties than possible in this work. 

 \acknowledgements

 J.A.G. acknowledges support from NASA grant
NNX15AV31G and from the Alexander von Humboldt
Foundation. R.M.T.C. has been supported by NASA
grant 80NSSC177K0515. VG is supported through the Margarete von Wrangell fellowship  by the ESF and the Ministry of Science, Research and the Arts  Baden-W\"urttemberg.
 
This research has made use of data, software and/or web tools obtained from the High Energy Astrophysics Science Archive Research Center (HEASARC), a service of the Astrophysics Science Division at NASA/GSFC and of the Smithsonian Astrophysical Observatory's High Energy Astrophysics Division. 

This research has made use of ISIS functions (ISISscripts) provided by 
ECAP/Remeis observatory and MIT (http://www.sternwarte.uni-erlangen.de/isis/).

\vspace{5mm}
\facilities{\rxte\ (PCA; \citealt{Jahoda1996}), HEASARC}

\software{{\tt XSPEC v.12.10.1c} \citep{Arnaud1996}, {\tt XILLVER} \citep{Garcia2010,Garcia2013}, {\tt RELXILL} (v1.2.0; \citealt{Garcia2014,Dauser2014}).}

\appendix
\section{{\tt pcacorr} validation}
\label{sec:app}
The results shown in Section~\ref{sec:refl_results} were dependent upon, to some degree, the calibration corrections provided by implementing the {\tt pcacorr} tool \citep{Garcia2014b}. Here we show that the model constraints were not skewed artificially by features which could be added during the reduction of systematic residuals, i.e., {\tt pcacorr} effectively removed narrow features and edge effects which allowed us to fit for source features at high counts, particularly in the soft X-ray band. 

Figure~\ref{fig:uncorr_tests} shows the ratio residuals of our fits to the corrected PCU~2 spectra. We show fits to observations 40401-01-27-00 (the very soft state), and 30188-06-03-00 (the hardest observation in our sample). The soft state spectrum is fit with the {\tt relxillNS} model, identically to the fit shown in Table~\ref{tab:soft_state}. The hard state fit is identical to that shown in Table~\ref{tab:params1}. For comparison, we have over-plotted the ratio residuals after replacing the corrected data with the uncorrected data (with no re-fitting). In the soft state, we see that due to the high number of X-ray counts in the $\sim3\mbox{--}6$~keV band, without applying {\tt pcacorr} the xenon L edge feature is far more pronounced, and also has narrow features. These were successfully removed when applying the {\tt pcacorr} tool, showing that the effect of applying the tool was to remove biases to the fit, as opposed to introducing them. We checked the parameter constraints which result from fitting the model to the uncorrected data (with identical systematics of $0.1\%$ applied to all channels), and see no changes. In the hard state, a similar effect is observed, albeit less pronounced, due to there being fewer counts in the soft band. We also tested for changes to model parameters when fitting to the uncorrected spectrum, and found slight differences. For example, the coronal electron temperature, $kT_{\rm e}$, decreases by a few keV, and $\Gamma$ increases by $\sim0.04$. These are changes which do not alter the parameter trends found in Section~\ref{sec:refl_results}. Thus application of {\tt pcacorr} reduces systematics which then allow more robust model fits to the PCA data, without skewing the physical interpretation of the results. 


\begin{figure*}
\centering
\includegraphics[width=0.4\linewidth]{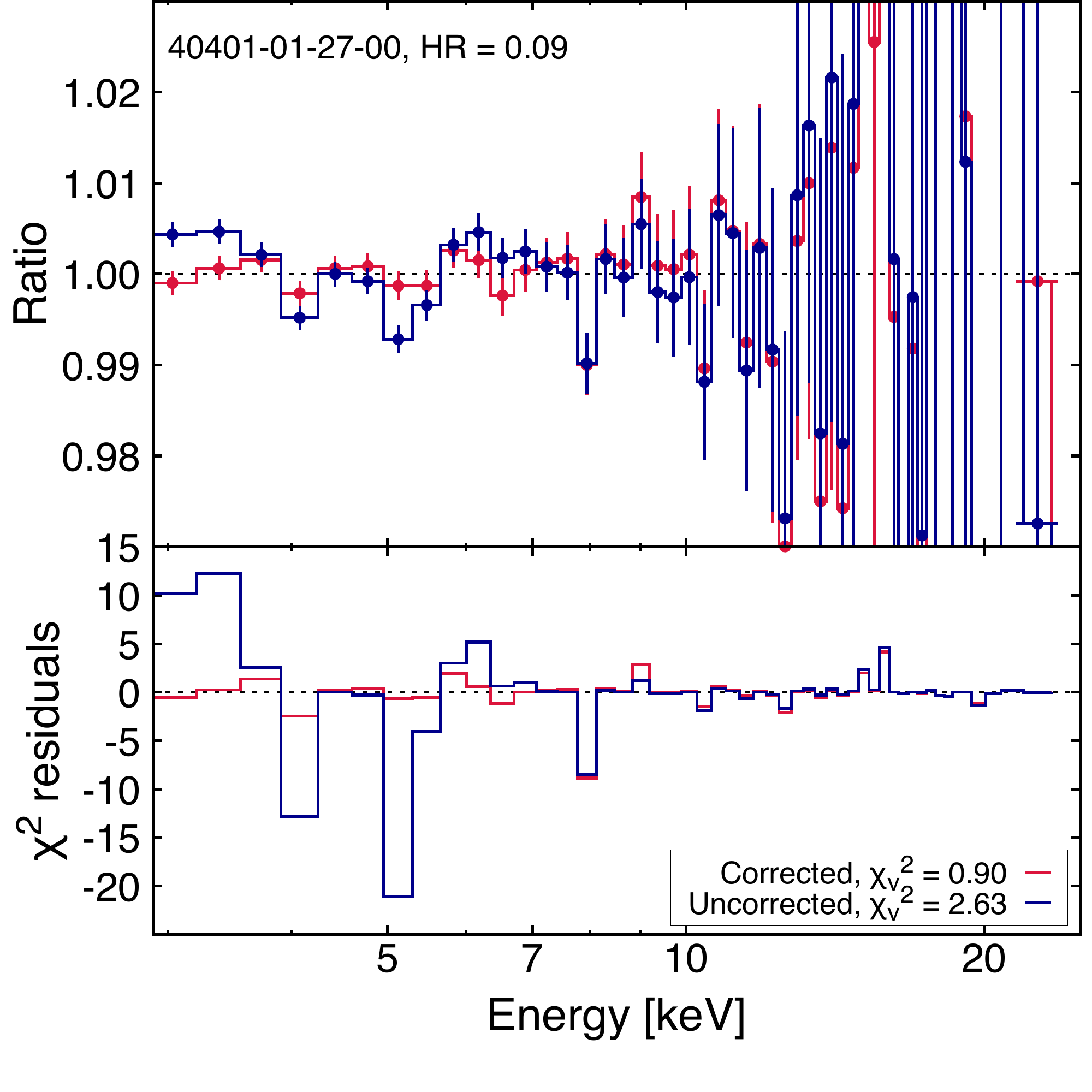}
\includegraphics[width=0.4\linewidth]{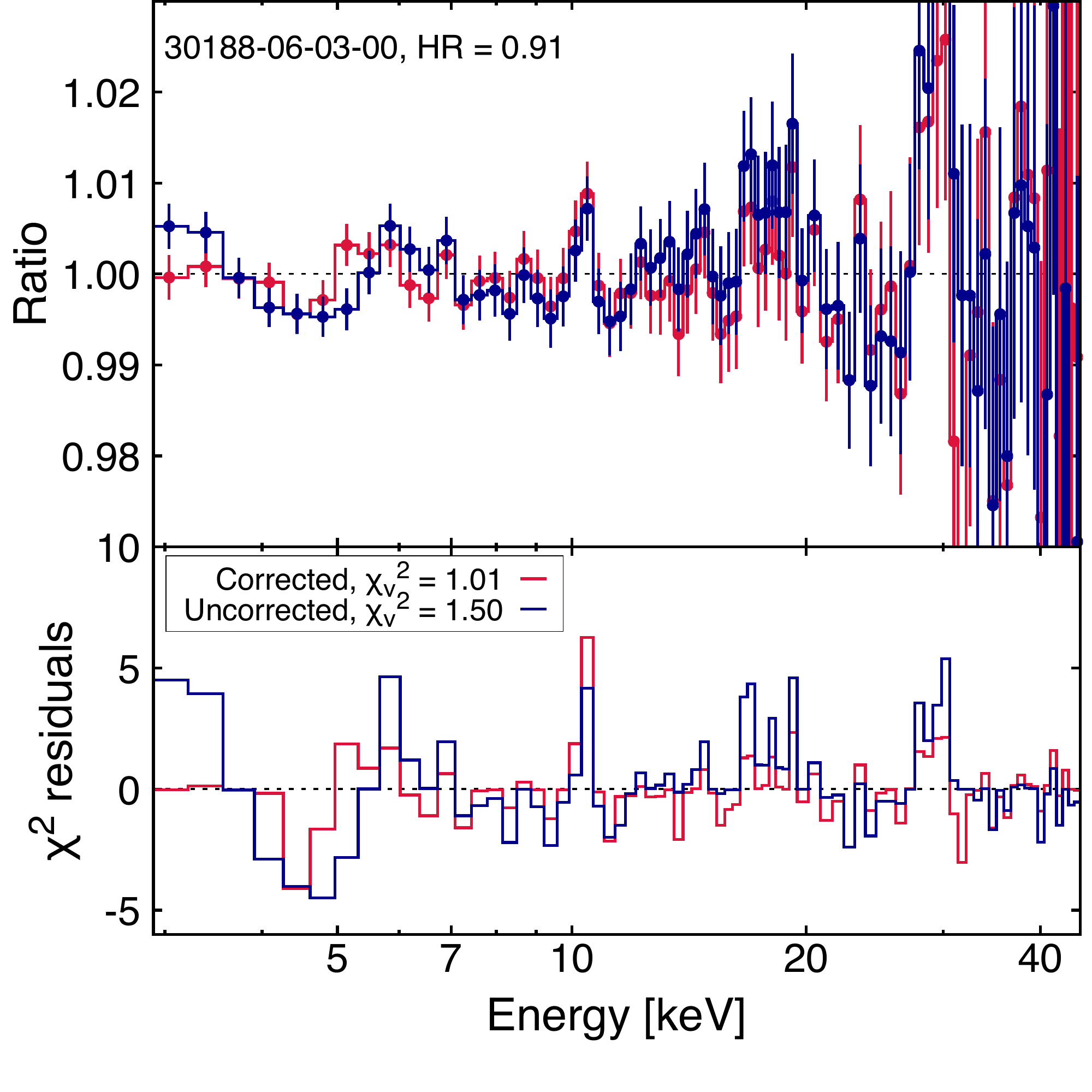}
\caption{Data/model ratio residuals of the best fit model to the soft and hard state data (red) and the uncorrected data (blue) for the same model. The difference in residuals quantifies the reduction in systematic features after applying the {\tt pcacorr} tool. In both the corrected and uncorrected PCA data, $0.1\%$ systematic errors are applied to all channels.}
\label{fig:uncorr_tests}
\end{figure*}

\section{The {\tt relxillNS} reflection model}
\label{sec:app_b}

Figure~\ref{fig:relxillNS} shows the differences between {\tt relxillNS} and {\tt relxillCp}. The spectral shapes are vastly different, as one expects when altering the irradiation spectrum from power-law-like, to a blackbody shape. In both the {\tt relxillNS} and {\tt relxillCp} model spectra, the default reflection parameters are set to $i=40^{\circ}$, $A_{\rm Fe}=1$, $\log\xi=3.1$, $a_{\star}=0.5$, $R_{\rm in}=R_{\rm ISCO}$. The {\tt relxillNS} model has a parameter for the disk density, which is fixed at $10^{15}~{\rm cm^{-3}}$, a standard value in these reflection models. The blackbody continuum in the {\tt relxillNS} model is given by $kT_{\rm b}=1.1$~keV. The IC continuum for the {\tt relxillCp} model is given by $\Gamma=2.5$, $kT_{\rm e}=300$~keV. For varying inclination, $i$, iron abundance, $A_{\rm Fe}$, and ionization, $\log\xi$, {\tt relxillNS} produces much softer, blackbody-like reflection, with significantly more curvature surrounding the Fe K region, and negligible flux beyond $\sim20$~keV. 

\begin{figure*}
\centering
\includegraphics[width=0.8\linewidth]{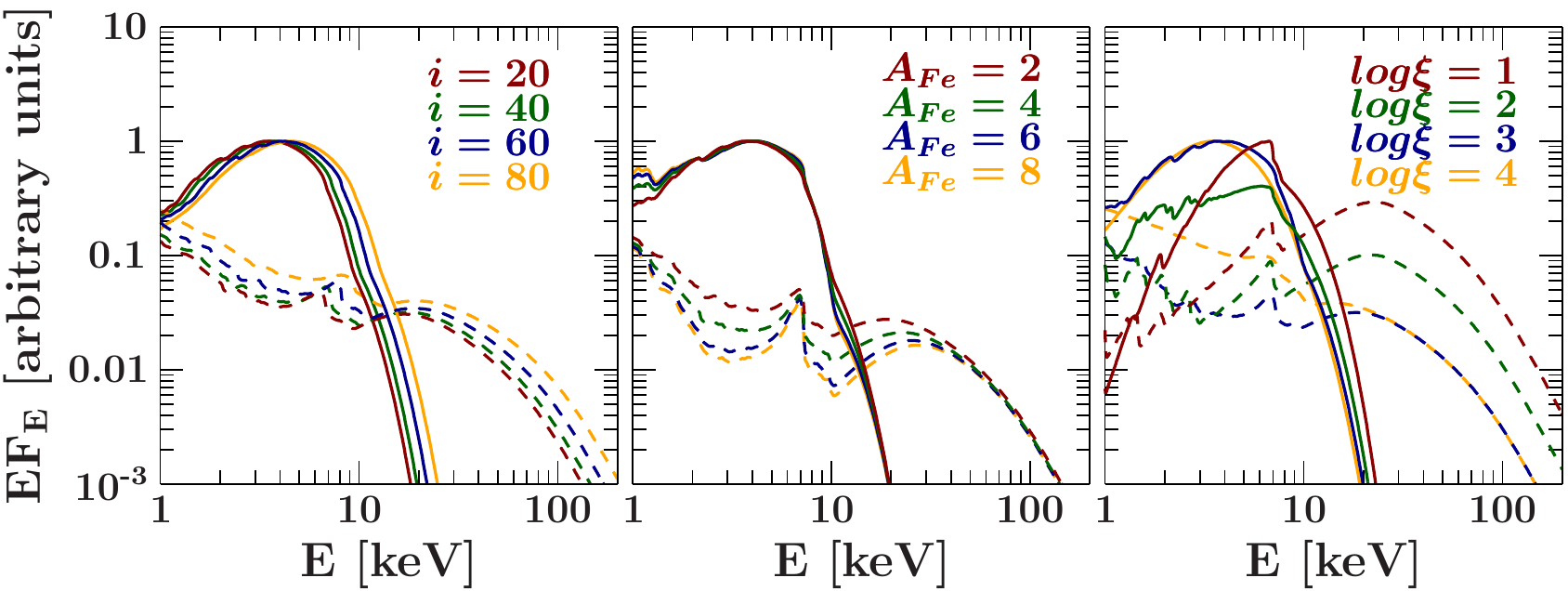}
\caption{A comparison of reflection spectra produced by the {\tt relxillNS} (solid lines) and {\tt relxillCp} (dashed lines) flavors of the {\tt relxill} quite of models. The {\tt relxillNS} model adopts a blackbody spectrum as the irradiating continuum, where {\tt relxillCp} adopts an IC spectrum. The three panels show variations in the models for different values of disk inclination, $i$, iron abundance, $A_{\rm Fe}$, and ionization, $\log\xi$. } 
\label{fig:relxillNS}
\end{figure*}

\section{The very soft state: testing different models}
\label{app_c}

As one additional clarifying test, we can ensure that no potential variations in the standard coronal IC flavor of reflection can successfully fit to the very soft state spectrum of \j1550. We invoked three extra model variations to check this: {\tt relxillD}, a high-density reflection model \citep{Garcia2016b}, {\tt relxillCp} with the emissivity index ($q$) free to vary, and {\tt relxilllpCp}, the lamppost flavor of {\tt relxillCp}. Here, instead of the emissivity profile being parameterized by the index $q$, a point source is located at a parameterized height, $h$, above the BH on the $z$ axis. Figure~\ref{fig:relxill_test} shows the $\chi^2$ residuals as a comparison of the fit quality of each of these models against the {\tt relxillNS} model already shown in Figure~\ref{fig:relxill_comp}. One can see that all three flavors of {\tt relxill} we tested suffer the same issues when fitting such a soft spectrum: they each overfit their Compton humps to the higher energies. It is clear that as long as the irradiating continuum is power-law-like, no alterations to either the emissivity or the disk properties can force such reflection models to fit to the soft state data.

\begin{figure}
\centering
\includegraphics[width=0.7\linewidth]{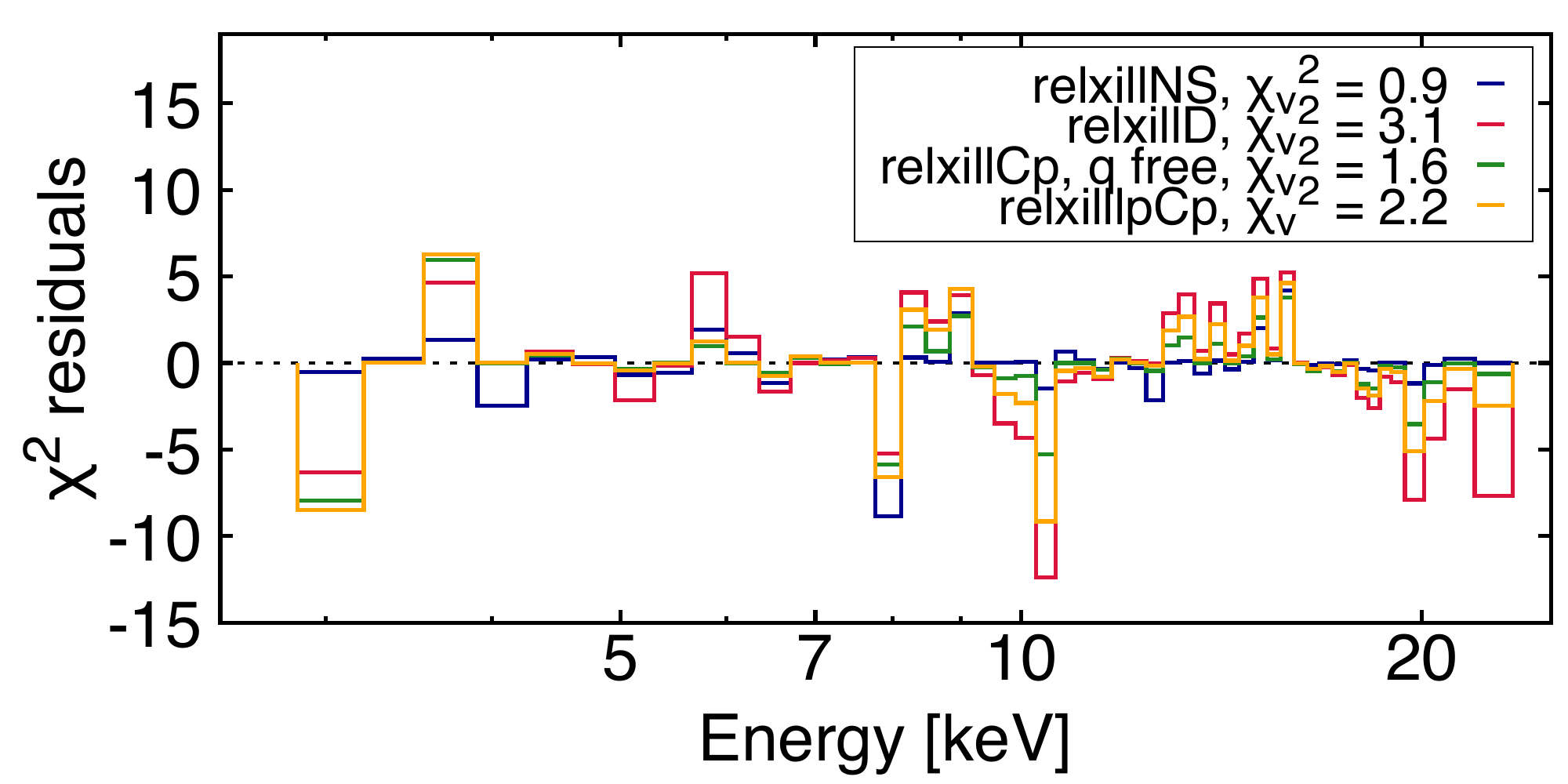}
\caption{$\chi^2$ residuals of fits to the soft state observation 40401-01-27-00 with different flavors of {\tt relxill}: {\tt relxillNS}, {\tt relxillD}, a high-density reflection model \citep{Garcia2016b}, {\tt relxillCp} with the emissivity index ($q$) free to vary, and {\tt relxilllpCp}, the lamppost flavor of {\tt relxillCp}, in which a point source lies at some height $h$ above the BH. The latter three models fit poorly to the soft state spectrum. }
\label{fig:relxill_test}
\end{figure}

\bibliographystyle{aasjournal}
\bibliography{references}
%
%
%
%

\end{document}